\documentclass{article}
\usepackage{amsmath,amssymb,amsfonts,latexsym,amscd,amsthm}
\usepackage{fancyhdr,color,epsfig,setspace,caption,times}
\usepackage[colorlinks=true,plainpages=false,linktocpage,linkcolor=blue,citecolor=green,pagebackref]{hyperref}
\usepackage[letterpaper,left=1.5in,top=1in,bottom=1in,right=1.5in,includeheadfoot,headheight=16.6pt]{geometry}
\usepackage{graphicx}

\begin{document}

\title{Playing Games with Quantum Mechanics}
\author{Simon J.D. Phoenix \&  Faisal Shah Khan \\ 
Khalifa University, PO Box 127788, Abu Dhabi, UAE}
\maketitle

\begin{abstract}
We present a perspective on quantum games that focuses on the physical aspects
of the quantities that are used to implement a game. If a game is to be
played, it has to be played with objects and actions that have some physical
existence. We call such games playable. By focusing on the notion of
playability for games we can more clearly see the distinction between
classical and quantum games and tackle the thorny issue of what it means to
quantize a game. The approach we take can more properly be thought of as
gaming the quantum rather than quantizing a game and we find that in this
perspective we can think of a complete quantum game, for a given set of
preferences, as representing a single family of quantum games with many
different playable versions. The versions of Quantum Prisoners Dilemma
presented in the literature can therefore be thought of specific instances of
the single family of Quantum Prisoner's Dilemma with respect to a particular
measurement. The conditions for equilibrium are given for playable quantum
games both in terms of expected outcomes and a geometric approach. We discuss
how any quantum game can be simulated with a classical game played with
classical coins as far as the strategy selections and expected outcomes are concerned.

\end{abstract}

\section{Playable Games}

Multiplayer non-cooperative game theory is a mathematical formulation of
competition in which players compete against one another to obtain some
resource or reward [1]. At its most abstract level a game is simply a mapping,
via some function, of the elements of one set to another. Although the
mathematics of game theory can be phrased in terms of such abstract symbols,
those symbols are given an interpretation in terms of the actions, or
strategies, of the players and the eventual rewards they receive. Implicit in
this interpretation is the notion that, should it be so desired, the players
could actually \textit{play} such a game. In other words, there exists a real
physical implementation of a game in terms of physically realizable actions
and tangible rewards. We call any game that can be implemented in a physical
reality a \textit{playable game}.

In a world described by classical physics the correspondence between the
mathematical abstractions and the real physical objects that might be used to
play a game is straightforward. Thus we might implement a game using counters
and the strategies would correspond to moves made with those counters. The
resultant state of the counters after the players' strategies have been
implemented determines the outcome of the game and we could imagine the
players receiving some tangible reward such as cash or cupcakes. Informally,
in such a scenario it is assumed that rational players will choose a strategy
that will maximise this quantity, given that the other players are choosing
their strategies to achieve the same maximisation of their reward.

The world as we know it, however, is governed by the laws of quantum mechanics
and classical properties emerge as a macroscopic limit to this more
fundamental description. In a quantum description it is not so straightforward
to assign elements of physical reality to the various components we might
choose for the implementation of a game [2]. Thus in a game played with
quantum objects obeying the laws of quantum mechanics it is not immediately
obvious how to draw the correspondence between the mathematical abstractions
of game theory and the various physical elements needed to implement a game.
By focusing carefully on the notion of \textit{playability} for games we
believe that these difficulties can be resolved. The perspective we develop
shows how the standard notions from game theory can be applied to games played
with quantum mechanical objects.

It is fair to say that there has been a mixed reaction to the merger of game
theory and quantum mechanics, first formally introduced over ten years ago
[3,4]. These original seminal papers of Meyer, and Eisert, Wilkens and
Lewenstein, however, have created an exciting and intriguing new area of
research. Possibly motivated by the recognition that the model of computation
based on Turing machines depends fundamentally on the nature of physical
reality [5], and the subsequent development of quantum algorithms [6], there
has perhaps been the hope that quantum mechanics will, somehow, lead to a
similar revolution in game theory. Such a revolution has not yet occurred, but
there have been indications of some tantalizing results. It is not uncommon to
hear the opinion that quantum mechanics has nothing to offer game theory and
vice versa. This pessimism may be due, in some part, to the various approaches
to terminology in which terms such as quantum games, quantized games, quantum
strategies and sometimes even quantized strategies, can be used in slightly
different ways. A major step forward in the clarification of the terminology
was taken by Bleiler [7] who introduced the term proper quantization to refer
to quantum games that are correct extensions of some underlying classical
game. Much work has focused on specific examples of games played with
entangled quantum systems (see, for example, [3,4, 8-18]) in which at first
sight it appears that in some circumstances quantum mechanics allows the
players to reach a more advantageous equilibrium, when compared to the
classical games upon which these quantum versions are based. Such comparisons
are, however, not straightforward and it is not always clear to what extent
the quantum extension can be said to be the correct quantum version of the
classical game that inspired it. By focusing here on \textit{playable} games
we adopt the approach that games are physical processes that are played with
objects and actions that have a physical existence.

In what follows we shall concentrate on 2-player games for convenience. It is
not difficult to extend the description to multiplayer non-cooperative games
with a greater number of players than 2.

\subsection{Classical Non-Cooperative Games}

In typical expositions of multiplayer non-cooperative game theory it is
assumed that the players each choose a strategy from some set of strategies
available to them. In a 2-player game with players \textit{A} and \textit{B}
we might denote the set of strategies available to player \textit{A} by
$S^{A}=\left\{  S_{1}^{A},S_{2}^{A},S_{3}^{A},S_{4}^{A},...\right\}  $ and
similarly for player \textit{B}. The strategies are thus no more than elements
of a set. The sets $S^{A}$ and $S^{B}$ do not need to be equal, nor do they
need to intersect. The choices the players make become the input to some
function that calculates their reward based on the individual choices they
have made. The rewards, or outcomes, for each player are expressed as a tuple
$(O_{j},O_{k})$ drawn from a set of outcome tuples $O$, the individual tuple
that is calculated being determined by the input strategy tuple. Thus, if the
players choice of strategy is described by $(S_{1}^{A},S_{2}^{B})$ and the
outcome tuple that is generated by this input is $(O_{3},O_{1})$ we say that
player \textit{A} receives the outcome $O_{3}$\ and player \textit{B} the
outcome $O_{1}$. If we assume that both players select their respective
strategies from the same set of possibilities then the function that takes the
input strategies to the outcomes can be described by the mapping $f:S\times
S\longrightarrow O$.

We still do not quite have a structure that can be described as a
\textit{game}. The purpose of the game is to compete, or in crude terms, to
win. There must be some outcomes from the set $O$ that \textit{A} prefers
above the others, and similarly for player \textit{B}. Thus we need to
describe a \textit{preference relation} over the outcomes for each player that
describes their desired outcomes in some order of preference. Player
\textit{A}, knowing the values of the function $f(S_{i},S_{j})$ will try to
ensure that his most desired outcome will occur, given that player \textit{B}
is doing exactly the same thing. It is clear that, depending on the preference
relations, both players may have to compromise and accept an outcome that is
not their most desired in order not to obtain an outcome that is less
preferable. If both players can select a strategy such that they would not
change it, irrespective of the choice of the other player, then this is an
equilibrium position known as a Nash equilibrium [19].

A Nash equilibrium is often not the most optimal for the players in the sense
that there can be a pair of strategy choices for \textit{A} and \textit{B}
which will increase their respective payoffs or rewards. A Pareto optimal
outcome is one in which there is no other outcome that gives at least the same
payoff for every player \textit{and} gives at least one player a better
outcome. Pareto optimization can be thought of as the outcome such that it
cannot be improved upon without reducing the payoff of at least one player.

Any 2-player game that can be described in the fashion we have just outlined
is a \textit{playable game}, provided that the function $f(S_{i},S_{j})$ is
\textit{computable}. \ By computable we mean that there is some algorithmic
procedure for determining the output for a given input. An excellent
discussion of this can be found in [20]. The strategy choices can be thought
of as no more than symbols from some alphabet. These symbols can be coded in
binary and transmitted to some Turing machine which takes the inputs and
computes the output which is the outcome tuple for the players. Thus we can
think of a 2-player game as a Turing machine in which there are two tape
inputs, one for player \textit{A} and one for player \textit{B,} in which each
player writes their selection of symbol, in binary, on the tape. The Turing
machine reads the inputs and computes the appropriate output in the form of an
outcome for each player. There are, however, two very important subtleties
that are easy to overlook in this classical description.

\subsubsection{The role of measurement in a playable classical game}

Implicit in the description of a playable classical game is the understanding
that the choices of the players can be determined, that is, \textit{measured}.
These become the inputs to the computable function that determines the
outcome, but in order to perform that computation, we require that the choices
can be distinguished. If we implement the game using some Turing machine then
that machine is implicitly assumed to be capable of reading the \textit{state}
of the tape. In other words, the state of the tape is \textit{measured}. This
becomes more transparent if we consider an implementation of our playable game
in terms of coins. Each player will prepare his coins in some state, specified
by a binary string that represents the chosen strategies. This state must be
read, or `decoded' in order to compute the outcomes. So, for example, if the
players are each given three coins then player \textit{A} may choose to
transmit the coins in the state \textit{HTH} which represents one of the 8
possible choices available to him. In order for the outcome to be correctly
computed, the state of these coins, and the state of the coins of player
\textit{B}, must be correctly measured.

\subsubsection{State preparation in a playable classical game}

Also implicit in the description of a playable classical game is the
understanding that the players perform some action, or set of actions, on an
\textit{initial state} in order to convey their choice. So if player
\textit{A} is given three coins in the state \textit{HHH} in the example above
then in order to achieve his desired output state he must perform the actions
$\overline{F}F\overline{F}$ where $F$ is `flip' and $\overline{F}$ is `don't
flip'. It is tempting to describe the 8 possible output choices of each player
in the 3-coin example as their possible choices of strategy. Equally, we could
describe the strategy choice as a combination of \textit{initial state} +
\textit{sequence of operations}, because in this case there is a one-to-one
correspondence. The operations, in this case to flip or not flip, are the
available actions of the players, that is, the things that the players do to
achieve their desired output state. It is more natural to think of what
actions the players take, \textit{given an initial state}, as a strategy
because this corresponds more closely with the physical situation. In other
words a strategy is a means of answering the question \textquotedblleft given
a start point, what must I do to achieve my desired end
point?\textquotedblright.

In order to illustrate the potential difficulty with the identification of an
output state as a strategy we shall consider the following \textit{playable}
classical game. Each player is given 3 coins prepared in the state
\textit{HHH}. These are placed in some device which performs the flip
operation, if so desired. Furthermore, each player is given the capability of
performing a flip on \textit{one} of the other player's coins. It is now no
longer possible to identify the output state of player \textit{A}'s coins as
his choice of strategy. It is certainly possible to construct a physical
device that would implement such a game, and thus the game is certainly
playable. The strategy set of player \textit{A} in this case is given by the
list of the options open to him, which includes his possible actions on the
coins of player \textit{B}. In this game each player has 32 possible choices
of action, that is, their strategy sets have 32 elements, giving 32$^{2}$
possible combinations of both players' strategies. The number of distinct
output states of the 6 coins is 64.

\bigskip

\bigskip

The example above shows that the detailed physical description of how we
actually implement a game, that is how we actually play the game, is critical,
because it determines the elements of our `strategy' set. The output state is
determined by the choice of what the players \textit{do}, given an initial
state, and it is this ouput state that is measured. The measurement result
becomes the input to the computable function that determines the outcomes for
the players. It is this detailed description of the physical implementation of
a game that is essential when we consider what it means to play a game using
quantum mechanical objects that obey the laws of quantum mechanics. It is this
detailed description that allows us to consider what it means to `quantize' a game.

\subsubsection{Modelling a playable classical game}

The careful examination of the elements, implicit and explicit, that are
required to implement a game in a physical reality leads us to a general model
of a playable game. This is shown in figure 1 in which these essential
elements are abstracted.

The initial state is simply the starting configuration of whatever physical
element is acted upon by the players in order to produce the output. In the
case of an implementation of the game using a Turing machine, for example, the
initial state might be just be a number of blank squares upon which the
players can act to produce their desired output. The players have a set of
actions they can perform upon this initial state in order to change its
configuration. In the case of the coins this is whether to flip or not. Thus
this step in the implementation of a playable game can be thought of as state
preparation. The combination of the initial state and the actions performed by
the players results in some output state. This is the state the players have
configured by their actions. The configuration of the output state must be
measured to produce a measurement result that captures the configuration of
the output state in some way. These results become the input to some function,
which can be thought of as a look-up table, that produces the outcomes or
payoffs for the players.

\begin{figure}
\centerline{\includegraphics[bb=1in .2in 9in 5in,scale=0.6]{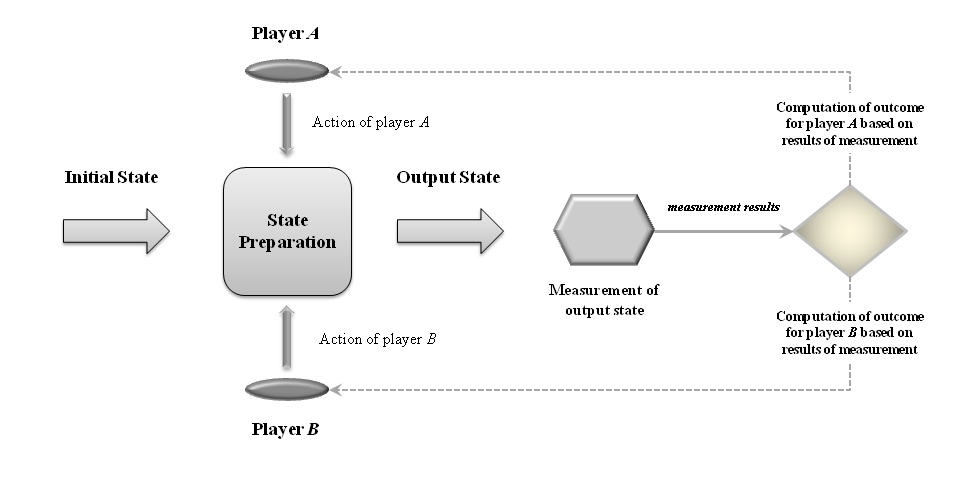}}
\caption{ Depiction of the essential elements of a playable classical game.}
\label{'PGWQM_Fig_1}
\end{figure}

The essential elements of a playable classical game, as described above,
suggest how we may approach the notion of quantizing a game. Strictly speaking
we quantize the physical system that is used to implement a classical game.
The question of what it means to `quantize' a game is somewhat problematical,
as we shall see. With this caveat in mind we now allow the physical elements
of our game to be quantum mechanical in nature with the operations of the
players being taken from the set of those permitted by quantum mechanics. The
measurement of the state becomes a quantum measurement. The results of the
measurement become the inputs to the computable function that determines the
outcomes for the players. It is important to note that the central
`philosophy' of the playable game has not altered by this extension to the
quantum domain; just as in the classical case, each player must consider what
operations he needs to perform on the initial state in order to produce an
output state the measurement of which will yield an optimal result for him,
given that the other player is making the same deliberation. Of course the
only real difference between games in the classical and quantum domains is
that in the latter the physical elements that implement the game are quantum
mechanical. The mathematical formalism of game theory applies in both domains.

\subsection{Quantum Games}

The description of a playable game implemented using classical objects implies
a certain underlying physical reality. According to the Copenhagen
interpretation, a quantum mechanical description of nature does not assign
elements of physical reality until a measurement is made. In classical game
theory it is tacitly assumed that that the choice of strategy corresponds to
some element of physical reality. In a Turing machine implementation of a
classical playable game, once a symbol to describe the strategy choice has
been written on the tape it is communicated to the Turing machine, and those
bits exist as elements of physical reality on the tape. In quantum mechanics,
however, these symbols, now have to be considered as qubits on a quantum
mechanical tape, and they do not have any corresponding element of physical
reality until they are measured. Furthermore, if we imagine two separate
inputs to our quantum Turing machine then not only does quantum mechanics mean
that we have to inscribe qubits on our quantum tape, but also that the two
quantum tape inputs can be \textit{entangled}.

Whilst the Turing machine model of a playable game is instructive in
highlighting the potential differences between playing games in the classical
and quantum domains, it is not the best model for allowing us to draw
correspondences between games played in the different domains. The general
model of a playable game given in Figure 1 is more suitable. The physical
elements of a playable game described in this figure remain unchanged when we
consider games in the quantum domain, the only difference being that the input
and output states are now quantum states and the actions the players perform
to transform the intial state to a desired output state are quantum operations.

In classical playable games, the players have some preference relation over
the measurement outcomes and it is these preference relations that determine
the choice of operation for the players. In the quantum domain the measurement
results are the eigenstates of the measurement operator and the players have
some preference relation over these eigenstates. (We only consider von Neumann
measurements in this paper. Extending to more generalized quantum measurements
adds unnecessary technical detail at present, and essentially introduces no
significant new conceptual elements to the game-theoretic description).

In order to be a little more specific we shall consider 2 player games in the
quantum domain. The extension to games with more than 2 players is relatively straightforward.

\subsubsection{2-Player Quantum Games}

We imagine a playable game with players \textit{A} and \textit{B} in which
some physical system, prepared in an initial quantum state $\left\vert
\psi_{0}\right\rangle $, is fed into a device. The players each perform some
unitary transformation on the state to produce an output state $\left\vert
\psi\right\rangle $. This output state is measured, or rather the physical
property $\hat{M}$ is measured, the result being one of the eigenstates,
$\left\vert \varphi_{j}\right\rangle $, of the operator $\hat{M}$.

Each player has some preference relation over these eigenstates which we label
$P_{A(B)}$. Before enacting their operation on the initial state, each player
performs some computation which determines which operation is in their best
interests, given that the other player is going to choose an operation which
is in \textit{his} best interests. The preference relations, of course,
determine what each player considers to be in their best interests given the
constraints. The choice of action is thus some computable function for each
player which takes as input the initial state, the possible measurement
outcomes, the preference relations and the available operations. Let us make
this more explicit.

We let the set of possible states of our quantum system be denoted by $\Psi$.
The set of all possible unitary operations on this state will be denoted by
$\hat{\Omega}$ where we have used the caret to remind ourselves that this is a
set of \textit{operators}. The ouput state from our device is thus described
by some mapping $g:\Psi\times\hat{\Omega}\times\hat{\Omega}\longrightarrow
\Psi$. The choice of the element from the set $\hat{\Omega}$\ for each player
is determined by their knowledge of the initial state, the operations that can
be performed on that state, the results of the measurement on the output state
and the preference relations over those results. It is important to note that
if both players have full knowledge of the parameters of the physical system,
including knowledge of each other's preference relations and available
operations, then each player can determine the element of $\hat{\Omega}$ that
the other player will pick, assuming rational players, and assuming that the
game admits the calculation of such a preferred choice.

We are now in a position to examine some general properties of playable
quantum games. These general considerations will be made more concrete when we
consider a simple, but highly non-trivial, example of a quantum game in
section 3.

\section{Playable Quantum Games : General Considerations}

It is clear that there is no fundamental difference between playing a game
with quantum mechanical objects and state preparation. Indeed, playing a
quantum game \textit{is} state preparation, with the prepared state being
determined by the preference relations over the measurement outcomes (along
with knowledge of the other important parameters, as discussed above). One
important question, from the perspective of game theory, is\ whether the game
parameters lead to the preparation of an \textit{equilibrium} output state by
rational players.

Before we look at the general structure of a quantum game we have proposed it
is worth noting here that we are considering here only those quantum games in
which the players are allowed to perform some \textit{unitary} operation on
the input state. Measurements are, of course, perfectly allowable quantum
operations. A more general treatment of quantum games would allow the players
to perform some measurement on the quantum state, possibly followed (or
preceded) by some unitary operation. We shall touch upon this briefly when we
consider uncertainty in quantum games shortly, but the main details of this
more general treatment will be published elsewhere.

The various parameters that characterise a game played using quantum
mechanical objects and operations are described below.\medskip

$\Psi$ : the set of all possible states of the physical system used to
implement the game\medskip

$\left\vert \psi_{0}\right\rangle $ : the initial state of the physical system
$(\left\vert \psi_{0}\right\rangle \in\Psi)\medskip$

$\Psi_{out}$ : the set of possible output states $(\Psi_{out}\subseteq
\Psi)\medskip$

$\left\vert \psi\right\rangle $ : the output state, that is, the state
produced by the operations of the players on the initial state. $\ (\left\vert
\psi\right\rangle \in\Psi_{out})\medskip$

$\hat{\Omega}$ : the set of all possible unitary transformations on an element
of $\Psi\medskip$

$\hat{\Omega}_{A(B)}$ : the set of all unitary operations available to player
\textit{A}(\textit{B}). $\hat{\Omega}_{A(B)}\subseteq\hat{\Omega}$ and
$\hat{\Omega}_{A}$ is not necessarily equal to $\hat{\Omega}_{B}$. Note that
these sets may contain the indentity operator which simply means that one
option that the players can choose is to do nothing to the initial
state.$\medskip$

$\hat{\alpha}_{i}$ : an element of the set $\hat{\Omega}_{A}$. $(\hat{\alpha
}_{i}\in\hat{\Omega}_{A})\medskip$

$\hat{\beta}_{j}$ : an element of the set $\hat{\Omega}_{B}$. $(\hat{\beta
}_{j}\in\hat{\Omega}_{B})\medskip$

$\hat{M}$ : the Hermitian operator describing the measurement on the output
state $\left\vert \psi\right\rangle $. The measurement produces the eigenstate
$\left\vert \varphi_{k}\right\rangle $ with probability $\left\vert
\left\langle \varphi_{k}|\psi\right\rangle \right\vert ^{2}\medskip$

$\left\vert \varphi_{k}\right\rangle $ : the eigenstates of $\hat{M}%
$.$\ (\left\vert \varphi_{k}\right\rangle \in\Psi)\medskip$

$P_{A(B)}$ : an ordered list of the eigenstates $\left\vert \varphi
_{k}\right\rangle $ such that the 1$^{st}$ element is \textit{A}(\textit{B})'s
most preferred measurement result, the 2$^{nd}$ element is the 2$^{nd}$ most
preferred result and so on. This is the preference relation for \textit{A}%
(\textit{B})\medskip

The basic sets are depicted in figure 2. If the players do not have access to
the full set of operations permitted by quantum mechanics on the initial state
then the output state will be a subset of the set of possible states of the
quantum system.

\begin{figure}
\centerline{\includegraphics[bb=1in .2in 9in 5in,scale=0.6]{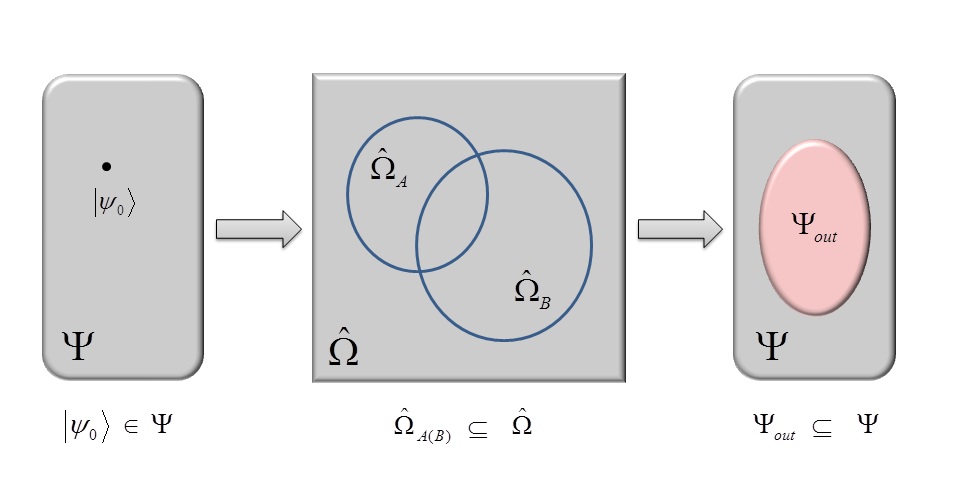}}
\caption{ The basic sets of a playable quantum game. An initial state is transformed by the choice of the
players' operation to produce an output state. The players' choice is
determined by their calculation of their expected outcome, given the
constraints of the other player's choice of operation.}
\label{'PGWQM_Fig_2}
\end{figure}

Despite the apparent increase in formalism introduced by the quantum
description the fundamental questions of game theory remain unaltered. What
actions must a player take to ensure that his most advantageous outcome
occurs, given that his opponent is making precisely the same deliberation? The
determination of what a particular player considers to be advantageous is, of
course, encapsulated in the preference relations. Another important
game-theoretic question that remains unaltered for a quantum game is whether
some equilibrium state is reached as a result of the players' deliberations.

Some immediate differences do, however, present themselves. The most obvious
of these is the difference between classical and quantum measurements. It is
not possible, in general, in quantum mechanics to determine the complete
configuration of a state. Indeed, it isn't even correct to ascribe elements of
reality to the elements of that configuration. In a classical game the
configuration of the output state can, in principle, be measured. In the
quantum case we can select a property of the state to measure, but this will
not give us complete information about the resultant output state.

This apparent limitation is partly overcome by relating the preference
relations of the players to the measurement outcomes. However, the consequence
of the quantum description is that some element of stochasticity is introduced
by measurement. In other words there will, in general, be a
\textit{probability distribution} over the outcomes in any game played using
quantum objects.

\subsection{State Preparation}

As we have discussed, a game played with quantum mechanical objects is nothing
more than state preparation followed by measurement in which the state that is
prepared depends upon the players' consideration of the possible measurement
outcomes, the input state and their available operations on that input state.
This leads to a certain fluidity in the description of a quantum game similar
to that when using the Heisenberg or Schr\"{o}dinger pictures.

Let us consider a quantum system that is used to play a game in which the
input state is $\left\vert \psi_{0}\right\rangle $, the sets of operations
available to the players are given by%

\begin{align*}
\hat{\Omega}_{A}  &  =\{\hat{\alpha}_{1},\hat{\alpha}_{2},...,\hat{\alpha}%
_{p}\}\\
\hat{\Omega}_{B}  &  =\{\hat{\beta}_{1},\hat{\beta}_{2},...,\hat{\beta}_{q}\}
\end{align*}

where we assume, for the moment that $\left[  \hat{\alpha}_{k},\hat{\beta}%
_{j}\right]  =0~~\forall j,k$. The output state is $\left\vert \psi
\right\rangle $, the measurement is characterized by the Hermitian operator
$\hat{M}$ which has eigenstates given by $(\left\vert \varphi\right\rangle
_{1},\left\vert \varphi\right\rangle _{2},...,\left\vert \varphi\right\rangle
_{k})$. Now let us suppose that we change the input state, keeping all other
parameters fixed. If the new input state is described by some transformation
of the initial state so that $\left\vert \psi_{0}^{\prime}\right\rangle
=T\left\vert \psi_{0}\right\rangle $ then, in general, a new output state
$\left\vert \psi^{\prime}\right\rangle $ will be produced. If we denote the
original game as game 1 and the game with the transformed input as game 2 then
the output states in the games are given by%

\begin{align*}
\text{game 1 }  &  \text{: }\left\vert \psi\right\rangle =\hat{\beta}_{j}%
\hat{\alpha}_{k}\left\vert \psi_{0}\right\rangle \\
\text{game 2 }  &  \text{: }\left\vert \psi^{\prime}\right\rangle =\hat{\beta
}_{j^{\prime}}\hat{\alpha}_{k^{\prime}}T\left\vert \psi_{0}\right\rangle
\end{align*}

where $\mathit{j,j}^{\prime},k$ and$\ k^{\prime}$ describe the operator
selections of the players. We can see that game 2 can be thought of as being
played with the input state $\left\vert \psi_{0}\right\rangle $ where player
\textit{A} has a different set of operators to select from given by%

\[
\hat{\Omega}_{A}^{\prime}=\{\hat{\alpha}_{1}T,\hat{\alpha}_{2}T,...,\hat
{\alpha}_{p}T\}
\]

We shall call this equivalent game, game 3. Alternatively we could view this
as a game played with an input $\left\vert \psi_{0}\right\rangle $ and the
operator sets%

\begin{align*}
\hat{\Omega}_{A}^{\prime\prime}  &  =\{T^{-1}\hat{\alpha}_{1}T,T^{-1}%
\hat{\alpha}_{2}T,...,T^{-1}\hat{\alpha}_{p}T\}\\
\hat{\Omega}_{B}^{\prime\prime}  &  =\{T^{-1}\hat{\beta}_{1}T,T^{-1}\hat
{\beta}_{2}T,...,T^{-1}\hat{\beta}_{q}T\}
\end{align*}

in which the state output by the players is then transformed by application of
$T$. If we call this new game, game 4, then we can see that games 2, 3 and 4
are all equivalent and they all result in a measurement of $\hat{M}$\ on the
state $\left\vert \psi^{\prime}\right\rangle $. We shall call a game for which
the output state is written in the form $\left\vert \psi\right\rangle
=\hat{\beta}\hat{\alpha}T\left\vert \psi_{0}\right\rangle $ an MW-type game. A
game for which the output state can be written in the form $\left\vert
\psi\right\rangle =T^{-1}\hat{\beta}\hat{\alpha}T\left\vert \psi
_{0}\right\rangle $ we shall call an EWL-type game. These are the forms of
game considered in the previous important work on quantum games [4,8] where
$T$ is an entanglement operator producing Bell basis states from the
computational basis. Because the calculations the players do are based on the
expected outcomes, the MW-type game can be thought of, somewhat informally, as
an EWL-type game in which the measurement operator is transformed (and vice
versa). In order to make a direct comparison between the two forms of games,
however, we should consider the same input state and the same measurement,
with the same preference relations over the eigenstates of the measurement operator.

\subsubsection{Relation between MW- and EWL-Type Games}

An MW-type game produces an output state of the form $\left\vert
\psi\right\rangle _{MW}=\hat{\beta}\hat{\alpha}T\left\vert \psi_{0}%
\right\rangle $ and an EWL-type game produces an output state of the form
$\left\vert \psi\right\rangle _{EWL}=T^{-1}\hat{\beta}\hat{\alpha}T\left\vert
\psi_{0}\right\rangle $. If we write $\left\vert \xi_{0}\right\rangle
=T\left\vert \psi_{0}\right\rangle $ then the MW-type game is nothing more
than the general form of quantum game we have introduced. The EWL-type is, in
a sense, more `artificial' because a subsequent rotation is imposed upon the
output state. If we use a fairly obvious notation for the operator sets so
that, for example, $\left(  \hat{\Omega}_{A}T\right)  =\{\hat{\alpha}%
_{1}T,\hat{\alpha}_{2}T,...,\hat{\alpha}_{p}T\}$ then the relation between the
forms of the MW- and EWL-type games can be seen in the following table for the
operator sets of the players where the games in each column are equivalent. We
have written the EWL-type game in this table from two perspectives that each
`eliminate' the rotation of the final state.\bigskip%

\begin{tabular}
[c]{ccc}
& MW-Type\medskip & EWL-Type\medskip\\
Input state : $\left\vert \xi_{0}\right\rangle \medskip$ & $\hat{\Omega}%
_{A},\hat{\Omega}_{B}$ & $\hat{\Omega}_{A},\left(  T^{-1}\hat{\Omega}%
_{B}\right)  $\\
Input state : $\left\vert \psi_{0}\right\rangle $ & $\ \ \ \left(  \hat
{\Omega}_{A}T\right)  ,\hat{\Omega}_{B}$ & $\ \ \ \ \ \ \ \ \ \ \left(
T^{-1}\hat{\Omega}_{A}T\right)  ,\left(  T^{-1}\hat{\Omega}_{B}T\right)  $%
\end{tabular}
\bigskip

Whether we view the operator sets as fixed or the input states as fixed it is
clear that the MW- and EWL-type games will in general lead to different output
states. The two EWL-type games in the table are equivalent to an EWL-type game
with input state $\left\vert \xi_{0}\right\rangle $, and operator sets
$\hat{\Omega}_{A}$ and $\hat{\Omega}_{B}$ in which a rotation is performed on
the resultant state. The rotation clearly alters the distribution of the
eigenstates of the measurement operator in the superposition. Of course both
players select their operation from their available set in the knowledge that
this final rotation will be performed. Alternatively we may view the EWL-type
game, as in the table, as one in which no rotation is performed but the
players have a transformed set of operators to select from.

As we shall see, it is an interesting question whether rational players would
prefer to play a game of the form $\left\vert \psi\right\rangle =\hat{\beta
}\hat{\alpha}\left\vert \psi_{0}\right\rangle ,~$ $\left\vert \psi
\right\rangle _{MW}=\hat{\beta}\hat{\alpha}T\left\vert \psi_{0}\right\rangle
,$ or $\left\vert \psi\right\rangle _{EWL}=T^{-1}\hat{\beta}\hat{\alpha
}T\left\vert \psi_{0}\right\rangle $ for a given $T$.

\subsection{Non-Commuting Games}

The next important element that a quantum mechanical description introduces is
the notion that operations on the input state do not necessarily commute.
Thus, in general, it matters whether player \textit{A} or player \textit{B}
performs their operation first. If, as above, we let the set of operations
available to player \textit{A}(\textit{B}) be denoted by $\hat{\Omega}_{A(B)}
$ where $\hat{\Omega}_{A(B)}\subseteq\hat{\Omega}$ and $\hat{\Omega}_{A}$ is
not necessarily equal to $\hat{\Omega}_{B}$ then a quantum game is
\textit{non-commuting} if there is at least one element $\hat{\alpha}_{i}%
\in\hat{\Omega}_{A}$ such that $\left[  \hat{\alpha}_{i},\hat{\beta}%
_{j}\right]  \neq0$ for at least one $j$ where $\hat{\beta}_{j}\in\hat{\Omega
}_{B}$.

Let us suppose that the players have access to a discrete set of operations
such that the cardinality of $\hat{\Omega}_{A}$ is \textit{p} and the
cardinality of $\hat{\Omega}_{B}$ is \textit{q. }If player \textit{A} makes
the first move then the possible output states are described by a $p\times q$
matrix with elements given by $\left\vert \psi_{ij}^{BA}\right\rangle
=\hat{\beta}_{j}\hat{\alpha}_{i}\left\vert \psi_{0}\right\rangle $. If player
\textit{B} makes the first move then the possible output states are described
by a $q\times p$ matrix with elements given by $\left\vert \psi_{ij}%
^{AB}\right\rangle =\hat{\alpha}_{j}\hat{\beta}_{i}\left\vert \psi
_{0}\right\rangle $. Thus if $\hat{\Omega}_{A}\neq\hat{\Omega}_{B}$ we have
that, in general, $\Psi_{out}^{BA}\neq\Psi_{out}^{AB}$.

Thus, in a non-commuting game a player's choice of operation may be determined
by whether he operates on the input state before or after the other player. Of
course, depending on the elements of the sets $\hat{\Omega}_{A(B)}$, it is
possible that a different equilibrium state is reached in the two situations.
Indeed, the choice of operation for each rational player will in general be
different depending on which player makes the first move. This immediately
raises the intriguing question of what happens if the players don't know which
of their operations is performed by the physical device first.

\subsubsection{Commutativity in Equivalent MW-Type Games}

As we have seen, there are different equivalent perspectives for an MW-type
game. We shall label these as MW$_{i}$ so that we have

\begin{itemize}
\item MW$_{0}$ : input state $\left\vert \xi_{0}\right\rangle =T\left\vert
\psi_{0}\right\rangle $, operator sets $\hat{\Omega}_{A},\hat{\Omega}_{B}$, no
final state rotation

\item MW$_{1}$ : input state $\left\vert \psi_{0}\right\rangle $, operator
sets $\left(  \hat{\Omega}_{A}T\right)  ,\hat{\Omega}_{B}$, no final state rotation

\item MW$_{2}$ : input state $\left\vert \psi_{0}\right\rangle $, operator
sets $\left(  T^{-1}\hat{\Omega}_{A}T\right)  ,\left(  T^{-1}\hat{\Omega}%
_{B}T\right)  $, final state rotation by $T$
\end{itemize}

Each of these perspectives leads to the production of the same output state
and therefore the expected outcomes for the players are unchanged in each of
these perspectives. We shall suppose that MW$_{0}$ is a commuting game so that
$\left[  \hat{\alpha}_{k},\hat{\beta}_{j}\right]  =0~~\forall j,k$.

Clearly MW$_{1}$ is a non-commuting game, in general, and it is only
equivalent if player \textit{A} plays first. The MW-type game with input
$\left\vert \psi_{0}\right\rangle $, operator sets $\hat{\Omega}_{A},\left(
\hat{\Omega}_{B}T\right)  $ in which player \textit{B} plays first, is also
equivalent to MW$_{0}$ when MW$_{0}$ is a commuting game. Accordingly we
denote these equivalent games as MW$_{1A}$\ and MW$_{1B}$\ where the
additional subscript denotes which player plays first.

If we denote the elements of the operator sets in MW$_{2}$ as $\hat{\alpha
}_{k}^{\prime}$ and $\hat{\beta}_{j}^{\prime}$ then the commutation relation
between them is given by%

\[
\left[  \hat{\alpha}_{k}^{\prime},\hat{\beta}_{j}^{\prime}\right]
=T^{-1}\left[  \hat{\alpha}_{k},\hat{\beta}_{j}\right]  T
\]

so that if MW$_{0}$ is a commuting game MW$_{2}$ is also a commuting game.

\subsubsection{Commutativity in Equivalent EWL-Type Games}

The different equivalent perspectives in EWL-type games are given by

\begin{itemize}
\item EWL$_{0A}$ : input state $\left\vert \xi_{0}\right\rangle =T\left\vert
\psi_{0}\right\rangle $, operator sets $\hat{\Omega}_{A},\left(  T^{-1}%
\hat{\Omega}_{B}\right)  $, no final state rotation, player \textit{A} plays first

\item EWL$_{0B}$ : input state $\left\vert \xi_{0}\right\rangle =T\left\vert
\psi_{0}\right\rangle $, operator sets $\left(  T^{-1}\hat{\Omega}_{A}\right)
,\hat{\Omega}_{B}$, no final state rotation, player \textit{B} plays first

\item EWL$_{1}$ : input state $\left\vert \psi_{0}\right\rangle $, operator
sets $\left(  T^{-1}\hat{\Omega}_{A}T\right)  ,\left(  T^{-1}\hat{\Omega}%
_{B}T\right)  $, no final state rotation, commuting game if $\left[
\hat{\alpha}_{k},\hat{\beta}_{j}\right]  =0~~\forall j,k$.

\item EWL$_{2A}$ : input state $\left\vert \psi_{0}\right\rangle $, operator
sets $\left(  \hat{\Omega}_{A}T\right)  ,\left(  T^{-1}\hat{\Omega}%
_{B}\right)  $, no final state rotation, player \textit{A} plays first

\item EWL$_{2B}$ : input state $\left\vert \psi_{0}\right\rangle $, operator
sets $\left(  T^{-1}\hat{\Omega}_{A}\right)  ,\left(  \hat{\Omega}%
_{B}T\right)  $, no final state rotation, player \textit{B} plays first

\item EWL$_{3A}$ : input state $\left\vert \psi_{0}\right\rangle $, operator
sets $\left(  \hat{\Omega}_{A}T\right)  ,\hat{\Omega}_{B}$, final state
rotation by $T^{-1}$, player \textit{A} plays first

\item EWL$_{3B}$ : input state $\left\vert \psi_{0}\right\rangle $, operator
sets $\hat{\Omega}_{A},\left(  \hat{\Omega}_{B}T\right)  $, final state
rotation by $T^{-1}$, player \textit{B} plays first
\end{itemize}

All these are different but equivalent perspectives leading to the production
of the same state upon which a measurement of $\hat{M}$ is performed, provided
that we have $\left[  \hat{\alpha}_{k},\hat{\beta}_{j}\right]  =0~~\forall
j,k$. As an example of the drastic difference that the order of play can make
for a non-commuting game consider the perspectives EWL$_{2A}$ and EWL$_{2B}$.
If we now swap the order of play in each of these perspectives we obtain the
commuting game with input $\left\vert \psi_{0}\right\rangle $ and operator
sets $\hat{\Omega}_{A}$ and $\hat{\Omega}_{B}$.

\subsection{Equilibrium in Quantum Games}

In game theory an equilibrium state is reached when the two players select a
strategy they would not change irrespective of the other player's choice. Such
a state is a Nash equilibrium. We can see from the general model of a playable
game that if such an equilibrium exists then it leads to a particular output
state. In quantum games, therefore, the existence of an equilibrium implies
that a particular output state is also produced (or possibly one of a family
of physically equivalent output states that yield the same distribution over
the measurement outcomes).

If we label the eigenstates of the measurement operator as $\left\vert
\varphi_{k}\right\rangle $ then the output state can be expanded in terms of
these eigenstates%

\[
\left\vert \psi\right\rangle =%
{\displaystyle\sum\limits_{k=1}^{n}}
\left\vert \varphi_{k}\right\rangle \left\langle \varphi_{k}|\psi
\right\rangle
\]

and the result of the measurement is the eigenstate $\left\vert \varphi
_{j}\right\rangle $ with probability $\left\vert \left\langle \varphi_{j}%
|\psi\right\rangle \right\vert ^{2}$. Intuitively, we might suppose that each
player will choose an operation that maximises the probability of their most
desired eigenstate according to their preference relations, \textit{given that
the other player is choosing his operation to achieve the same end }according
to his preference relations. Thus we can imagine that there are two competing
`forces' trying to rotate the initial state to some preferred direction. This
geometric approach, briefly outlined below, has been explored elsewhere [21,22].

In general the measurement of the output state will lead to a distribution of
outcomes and consequently the players must seek to optimise their
\textit{expected} outcomes. Let us suppose that player \textit{A} performs his
operation first. Player \textit{A} will choose an $\hat{\alpha}_{i}$ so that
his expected outcome is optimised whatever \textit{B} subsequently does.
Likewise, \textit{B} will choose a $\hat{\beta}_{j}$ so that his expected
outcome is optimised whatever \textit{A} has chosen.

Let us consider a measurement $\hat{M}$ with \textit{n} eigenstates given by
the list $(\left\vert \varphi_{1}\right\rangle ,\left\vert \varphi
_{2}\right\rangle ,...,\left\vert \varphi_{n}\right\rangle )$ which for
convenience we can write as the list of numbers $\varphi=(1,2,..,n)$. The
preference relation of player \textit{A} is simply a permutation of this list
such that the first state in the list is his most preferred, the second state
his second most preferred, and so on, and similarly for player \textit{B}. If
we let $\pi^{A(B)}(k)$ be functions that take the integers from 1 to
\textit{n} as input and output the number from $\varphi$\ that represents the
\textit{k}$^{th}$ most preferred state of player \textit{A}(\textit{B}) then
we can write the preference relations for the players as an ordered
\textit{n}-tuple of the integers from 1 to \textit{n}.%

\begin{align*}
P_{A}  &  =(\pi^{A}(1),\pi^{A}(2),...,\pi^{A}(n))\\
P_{B}  &  =(\pi^{B}(1),\pi^{B}(2),...,\pi^{B}(n))
\end{align*}

In order to calculate an expected \textit{outcome} we must assign some
numerical value, or weight, to each of the eigenstates that encapsulates the
notion of preference. In crude terms this numerical value can be thought of as
the payoff for each state. For simplicity we attach the numerical weight
\textit{n} to the most preferred state, and the weight 1 to the least
preferred state, for each player. Thus if the measurement result yields
\textit{A}'s most preferred state then player \textit{A} would receive
\textit{n} cupcakes, for example. The expected payoffs for the players,
denoted by $E_{ij}^{A(B)}$, if the operations $\hat{\alpha}_{i}$ and
$\hat{\beta}_{j}$ are chosen, are then given by%

\begin{align*}
E_{ij}^{A}  &  =\sum_{k=1}^{n}(n-k+1)\left\vert \left\langle \varphi_{\pi
^{A}(k)}|\hat{\beta}_{j}\hat{\alpha}_{i}|\psi_{0}\right\rangle \right\vert
^{2}\\
E_{ij}^{B}  &  =\sum_{k=1}^{n}(n-k+1)\left\vert \left\langle \varphi_{\pi
^{B}(k)}|\hat{\beta}_{j}\hat{\alpha}_{i}|\psi_{0}\right\rangle \right\vert
^{2}%
\end{align*}

These quantities define a matrix of expected outcomes for each player, the
elements denoting each choice of operation $(\hat{\alpha}_{i},\hat{\beta}%
_{j})$. We have assumed that player \textit{A} plays first and the knowledge
of the order of the moves can make a difference to the strategy selection. In
a non-commuting game the order of play must be specified. For a commuting game
in which there is no specification of the order of play then if \textit{A} and
\textit{B} can find operators $(\hat{\alpha}_{i},\hat{\beta}_{j})$\ such that%

\begin{align*}
\sum_{j}E_{ij}^{A}  &  >\sum_{j}E_{pj}^{A}~~~~(\forall p\neq i)\\
\sum_{i}E_{ij}^{B}  &  >\sum_{i}E_{iq}^{B}~~~~(\forall q\neq j)
\end{align*}

then an equilibrium output state is obtained given by $\left\vert
\psi\right\rangle =\hat{\beta}_{j}\hat{\alpha}_{i}\left\vert \psi
_{0}\right\rangle =\hat{\alpha}_{i}\hat{\beta}_{j}\left\vert \psi
_{0}\right\rangle $. The players thus each maximise their expected outcomes
over the choices of the other player.

For a non-commuting game in which player \textit{A} plays first, the strategy
selection is influenced by the order of play. Player \textit{A} must select
his strategy knowing that \textit{B} has the advantage of playing second.
Player \textit{A} examines the matrix of expected outcomes for player
\textit{B} and looks at the choices of \textit{B} that maximise \textit{B}'s
expected outcome for every choice of player \textit{A}'s operator. This gives
\textit{A} a list of operator pairs. From these \textit{A} now selects the one
that maximises his expected outcome.

\subsubsection{The geometric approach to equilibrium}

When playing a game with quantum objects the players have some preference over
the measurement outcomes which are the eigenstates of the measurement
operator. We have written these above simply as an ordered list of the labels
for the eigenstates. More explicitly we can write these preferences as%

\begin{align*}
\text{Player \textit{A} }  &  \text{:}|\varphi_{\pi^{A}(1)}>~\succ
~|\varphi_{\pi^{A}(2)}>~\succ...\succ~|\varphi_{\pi^{A}(n)}>\\
\text{Player \textit{B} }  &  \text{:}|\varphi_{\pi^{B}(1)}>~\succ
~|\varphi_{\pi^{B}(2)}>~\succ...\succ~|\varphi_{\pi^{B}(n)}>
\end{align*}

Any output state from the game, before measurement, can be described by%

\[
\left\vert \psi\right\rangle =\sum_{k=1}^{n}\left\vert \varphi_{\pi^{A}%
(k)}\right\rangle \left\langle \varphi_{\pi^{A}(k)}|\psi\right\rangle
=\sum_{k=1}^{n}\left\vert \varphi_{\pi^{B}(k)}\right\rangle \left\langle
\varphi_{\pi^{B}(k)}|\psi\right\rangle
\]

Let us consider the subsets $\Psi_{\pi^{A(B)}}$ of $\Psi$\ given by the states%

\[
\left\vert \psi(\pi^{A(B)})\right\rangle =\left\langle \varphi_{\pi^{A(B)}%
(1)}|\psi\right\rangle \left\vert \varphi_{\pi^{A(B)}(1)}\right\rangle
+\left\langle \varphi_{\pi^{A(B)}(2)}|\psi\right\rangle \left\vert
\varphi_{\pi^{A(B)}(2)}\right\rangle +...+\left\langle \varphi_{\pi^{A(B)}%
(n)}|\psi\right\rangle \left\vert \varphi_{\pi^{A(B)}(n)}\right\rangle
\]

where the probability amplitudes respect the preferences so that%

\[
\left\vert <\varphi_{\pi^{A(B)}(1)}|\psi>\right\vert ^{2}~>~\left\vert
<\varphi_{\pi^{A(B)}(2)}|\psi>\right\vert ^{2}~>...>~\left\vert <\varphi
_{\pi^{A(B)}(n)}|\psi>\right\vert ^{2}%
\]

The subsets $\Psi_{\pi^{A(B)}}$ define regions of the full Hilbert space of
the outputs. These subsets are not subspaces and the elements of $\Psi
_{\pi^{A}}$ are not, in general, orthogonal to those of $\Psi_{\pi^{B}}$. Each
player would act in such a way as to try to direct the output state into their
respective regions of Hilbert space. The final output state will, therefore,
be some state that would be `as equidistant' from their respective regions
that the players can achieve with their given operator sets. The resultant
state, if this can be achieved, will be the Nash equilibrium position for the
quantum game. Some ramifactions of this approach are discussed in [21,22].

\subsection{The Preference Relations}

Let us suppose, as above, that the players have access to a discrete set of
operations such that the cardinality of $\hat{\Omega}_{A}$ is \textit{p} and
the cardinality of $\hat{\Omega}_{B}$ is \textit{q. }If player \textit{A}
makes the first move then the possible output states are described by an
$p\times q$ matrix with elements given by $\left\vert \psi_{ij}^{BA}%
\right\rangle =\hat{\beta}_{j}\hat{\alpha}_{i}\left\vert \psi_{0}\right\rangle
$. The players each have a preference relation over the eigenstates
$\left\vert \varphi_{k}\right\rangle $ of the measurement operator $\hat{M}$.
If \textit{n} is the cardinality of the set of eigenstates then, in general,
$n\neq pq$. There will often be a greater number of possible output states
that can be produced by the players than the number of possible measurement
outcomes. Indeed, as we shall see, the possible ouput states can form a
continuum whereas the measurement outcomes consist of a small discrete set of possibilities.

As we have discussed in the previous section, each player can determine a
matrix of expected outcomes where the elements $E_{ij}^{A(B)}$ give the
expected outcome for player $A(B)$ for the possible output state $\left\vert
\psi_{ij}\right\rangle =\hat{\beta}_{j}\hat{\alpha}_{i}\left\vert \psi
_{0}\right\rangle $. Clearly, this leads to an ordered list of expected
outcomes in which each player ranks each possible output state according to
its expected outcome for him. Thus the preference relation over the
eigenstates of the measurement operator for player $A(B)$ \textit{induces} a
preference relation over the possible output states.

Let us suppose that the measurement operator for some 2-player quantum game is
rotated by the action of an operator $T$ such that the new meaurement operator
is $\hat{M}^{\prime}=T^{-1}\hat{M}T$. The eigenstates of $\hat{M}^{\prime}$
are $\left\vert \varphi_{k}^{\prime}\right\rangle =T\left\vert \varphi
_{k}\right\rangle $. If the players maintain the same preference relation over
the new eigenstates so that $\left(  \pi^{A}(k)\right)  ^{\prime}=\pi^{A}(k)$
and $\left(  \pi^{B}(k)\right)  ^{\prime}=\pi^{B}(k) $ then, in order for the
players to be playing the same game with respect to the new measurement basis
the various components have to be transformed according to%

\begin{align*}
\left\vert \psi_{0}^{\prime}\right\rangle  &  =T\left\vert \psi_{0}%
\right\rangle \\
\hat{\alpha}_{i}^{\prime}  &  =T\hat{\alpha}_{i}T^{-1}\\
\hat{\beta}_{j}^{\prime}  &  =T\hat{\beta}_{j}T^{-1}%
\end{align*}

With these transformations the expected outcomes for player \textit{A} with
respect to the new measurement basis are given by%

\[
\left(  E_{ij}^{A}\right)  ^{\prime}=\sum_{k=1}^{n}(n-k+1)\left\vert
\left\langle \varphi_{\pi^{A}(k)}^{\prime}|\hat{\beta}_{j}^{\prime}\hat
{\alpha}_{i}^{\prime}|\psi_{0}^{\prime}\right\rangle \right\vert ^{2}%
=\sum_{k=1}^{n}(n-k+1)\left\vert \left\langle \varphi_{\pi^{A}(k)}|\hat{\beta
}_{j}\hat{\alpha}_{i}|\psi_{0}\right\rangle \right\vert ^{2}=E_{ij}^{A}%
\]

and similarly for player \textit{B}. Thus the expected outcomes remain
unaltered, provided that both the initial state and the operator choices of
the players are transformed.

\subsection{Invertible Games}

We define an invertible game as one in which one, or both, players can invert
the operations performed by the other player. For example, let us suppose that
the set of operations available to \textit{A} is given by $\{\hat{\alpha}%
_{1},\hat{\alpha}_{2},...,\hat{\alpha}_{p}\}$ then the game is invertible if
\textit{B} has a set of available operations that includes the inverses of
these elements, that is, the set of operations available to \textit{B} is
given by $\{\hat{\beta}_{1},\hat{\beta}_{2},...,\hat{\beta}_{q},\hat{\alpha
}_{1}^{-1},\hat{\alpha}_{2}^{-1},...,\hat{\alpha}_{p}^{-1}\}$. There is
nothing intrinsically quantum mechanical about such games for we could
construct a playable classical game in which the operations of one, or both,
players could be inverted by the other.

It is clear that an equilibrium state, in the sense that an equilibrium state
is the optimal choice for both players, does not exist for an invertible game.
In order to see this we shall consider a game in which \textit{A} moves first.
If we assume rational players then \textit{A}'s preferred choice of move given
the constraints, if it exists, is computable by both \textit{A} and
\textit{B}. Therefore \textit{B} simply needs to follow \textit{A}'s move with
$\hat{\beta}_{k}\hat{\alpha}^{-1}$ where $\hat{\beta}_{k}$ operating on
$\left\vert \psi_{0}\right\rangle $ maximises \textit{B}'s expected outcome.
Of course, this implies that there is no such rational choice for \textit{A}.
In fact, \textit{A}'s best strategy in this situation is to select his move
with some distribution.

One immediate consequence of this is that if we allow players to access every
possible operation on the input state allowed by quantum mechanics then there
cannot be an equilibrium output state. In order to reach some kind of
equilibrium we must restrict the available strategies in some way. This, of
course, restricts the possible output states. This consideration becomes
crucial when we consider our simple, but important, example of a quantum game.

If we suppose that $\hat{\Omega}_{A}=\hat{\Omega}_{B}$ and that the elements
of these sets form a group then the resulting game will be invertible. For
games which satisfy this condition there can be no equilibrium state.

\subsection{Factorable Quantum Games}

We define a factorable game as one in which there is no element of
entanglement. In other words, the input states are tensor product states and
the available operations only lead to output states that are also tensor
product states. Factorable games can still be non-commuting and invertible.

\subsection{Sequential Quantum Games}

A sequential game is one in which the players make a sequence of moves. So,
for example, player \textit{A} goes first, followed by player \textit{B}.
Player \textit{A} then makes his second move, followed by player \textit{B},
and so on. If we let $\hat{\alpha}(i)$ be the \textit{i}$^{th}$ operation of
player \textit{B} and $\hat{\beta}(i)$ be the \textit{i}$^{th}$ operation of
player \textit{B} then the output state after \textit{m} moves is given by%

\[
\left\vert \psi\right\rangle =\hat{\beta}(m)\hat{\alpha}(m)...\hat{\beta
}(2)\hat{\alpha}(2)\hat{\beta}(1)\hat{\alpha}(1)\left\vert \psi_{0}%
\right\rangle
\]

After each move we can think of the resultant state as being equivalent to
some new input state, thus a sequential game is a sequence of single-move
games in which the output state from the previous game becomes the input state
to the next. If we let $\left\vert \psi_{0}^{k}\right\rangle $ be the state
after the \textit{k}$^{th}$ move of the players then any sequential game of
\textit{m} moves can be thought of as a single-move game in which the input
state is given by $\left\vert \psi_{0}^{m-1}\right\rangle $ with%

\[
\left\vert \psi_{0}^{m-1}\right\rangle =\hat{\beta}(m-1)\hat{\alpha
}(m-1)...\hat{\beta}(2)\hat{\alpha}(2)\hat{\beta}(1)\hat{\alpha}(1)\left\vert
\psi_{0}\right\rangle
\]

Alternatively we can factor the sequence of operations into two new operators
$\hat{\alpha}^{\prime}$ and $\hat{\beta}^{\prime}$ where, for example,%

\begin{align*}
\hat{\alpha}^{\prime}  &  =\hat{\beta}(k)\hat{\alpha}(k)...\hat{\beta}%
(2)\hat{\alpha}(2)\hat{\beta}(1)\hat{\alpha}(1)\\
\hat{\beta}^{\prime}  &  =\hat{\beta}(m)\hat{\alpha}(m)...\hat{\beta}%
(k+2)\hat{\alpha}(k+2)\hat{\beta}(k+1)\hat{\alpha}(k+1)
\end{align*}

Thus the sequential game of \textit{m} moves is equivalent to a single-move
game in which the operators $\hat{\alpha}^{\prime}$ and $\hat{\beta}^{\prime}$
are included in \textit{A} and \textit{B}'s set of available operators,
respectively. Note that there is no unique way to perform this factorization.

These considerations also show us that any single-move game can be represented
by some equivalent sequential game of \textit{m} moves. Of course there is no
unique \textit{m}-move representation. In fact there are an infinite number of
\textit{m}-move representations of any single-move game. By equivalent we mean
that the same output states are produced and the players receive the same
expected outcomes. In game theory terms the players are in effect playing a
different game. Constructing the equivalent representations of a single-move
game or an \textit{m}-move game is not, in general, a trivial exercise.

If a single-move game has an equilibrium then all equivalent sequential
\textit{m}-move games also reach an equilibrium. Alternatively if we construct
an equivalent sequential \textit{m}-move game that reaches equilibrium then
the single-move game also reaches equilibrium. Indeed, if it can be shown that
one member of the class of equivalent \textit{m}-move games reaches
equilibrium then all games in the class also reach equilibrium, as does the
equivalent single-move game.

\subsection{Equilibrium as a Limit of a Sequential Game}

Another way to think of the production of an equilibrium state is as the limit
of a sequential game. Let us suppose that two players play a sequential game
with a no limit on the number of moves they can make. If they both reach a
point such that they do not wish to make another move after a finite number of
moves then the resultant output state is an equilibrium state. Conversely, if
no such limit exists then there cannot be an equilibrium.

If we consider an invertible sequential game with no fixed number of moves
then we can see that the players would never reach a point where they would
wish to stop. If we extend a single-move game to a sequential game with no
fixed number of moves then the resultant game may not have an equilibrium
state even if the single-move game reaches equilibrium. This is because the
subsequent moves potentially give the players access to a larger set of
operations than those for the single-move game. In order to see this let us
consider a sequential game with just 2 moves allowed. The output state is
given by%

\[
\left\vert \psi\right\rangle =\hat{\beta}(2)\hat{\alpha}(2)\hat{\beta}%
(1)\hat{\alpha}(1)\left\vert \psi_{0}\right\rangle
\]

Thus, in effect, \textit{A} chooses the operator $\hat{\alpha}^{\prime}%
=\hat{\alpha}(2)\hat{\beta}(1)\hat{\alpha}(1)$ which may not be included in
his set of available operations for his first move. We could also view this as
a single-move game in which \textit{A} plays $\hat{\alpha}(1)$ and \textit{B}
plays $\hat{\beta}^{\prime}=\hat{\beta}(2)\hat{\alpha}(2)\hat{\beta}(1)$ Thus
if the original single-move game has initial state $\left\vert \psi
_{0}\right\rangle $ and sets $\hat{\Omega}_{A}=\{\hat{\alpha}_{1},\hat{\alpha
}_{2},...,\hat{\alpha}_{p}\}$ and $\hat{\Omega}_{B}=\{\hat{\beta}_{1}%
,\hat{\beta}_{2},...,\hat{\beta}_{q}\}$ then the sequential game with 2 moves
is equivalent to the single-move game with initial state $\left\vert \psi
_{0}\right\rangle $ and sets $\hat{\Omega}_{A}^{\prime}=\{\hat{\alpha}%
_{1},\hat{\alpha}_{2},...,\hat{\alpha}_{p},\hat{\alpha}^{\prime}\}$ and
$\hat{\Omega}_{B}=\{\hat{\beta}_{1},\hat{\beta}_{2},...,\hat{\beta}_{q}\}$. It
is also equivalent to the single-move game with initial state $\left\vert
\psi_{0}\right\rangle $ and sets $\hat{\Omega}_{A}=\{\hat{\alpha}_{1}%
,\hat{\alpha}_{2},...,\hat{\alpha}_{p}\}$ and $\hat{\Omega}_{B}^{\prime
}=\{\hat{\beta}_{1},\hat{\beta}_{2},...,\hat{\beta}_{q},\hat{\beta}^{\prime
}\}$.

The 2-move sequential game also gives us a way of characterizing equilibrium
for a single-move game. If the players in a 2-move game would choose the
identity operation from their set of available operators for their second
move, this implies that their first move was the best they could make and they
cannot improve upon it by playing another. Thus equilibrium is reached for the
single-move game if the second move of a 2-move sequential game is the
identity operation, for both players. Equilibrium for an \textit{m}-move game
can also be defined in a similar fashion; if the $(m+1)^{th}$ move is the
identity element for both players, then the game has reached equilibrium.

\subsection{Quantum Games and Uncertainty}

We have seen one way in which considerations of non-commutativity impact upon
quantum games. Let us now look at another way in which this can arise within
the context of a quantum game. Consider the situation shown in figure 3 where
we have 2 quantum games. In both games the same input state $\left\vert
\psi_{0}\right\rangle $\ is used and the players have access to the same
operations $\hat{\Omega}_{A}=\{\hat{\alpha}_{1},\hat{\alpha}_{2}%
,...,\hat{\alpha}_{p}\}$ and $\hat{\Omega}_{B}=\{\hat{\beta}_{1},\hat{\beta
}_{2},...,\hat{\beta}_{q}\}$. The games differ, therefore, only in the
measurement of the output state. In game 1 the measurement performed is
$\hat{M}$ and in game 2 the measurement is $\hat{N}$.

Adopting a slightly different notation to that used previously, the
eigenvalues and eigenstates of these operators are given by%

\begin{align*}
\hat{M}\left\vert \mu_{k}\right\rangle  &  =\mu_{k}\left\vert \mu
_{k}\right\rangle \\
\hat{N}\left\vert \nu_{k}\right\rangle  &  =\nu_{k}\left\vert \nu
_{k}\right\rangle
\end{align*}

and we assume that $\left[  \hat{M},\hat{N}\right]  \neq0$. The set of
possible output states in both games is the same. The preference relations in
game 1 are given by%

\begin{align*}
P_{A}^{1}  &  =\left(  \left\vert \mu_{\pi^{A}(1)}\right\rangle ,\left\vert
\mu_{\pi^{A}(2)}\right\rangle ,...,\left\vert \mu_{\pi^{A}(n)}\right\rangle
\right)  \equiv(\pi^{A}(1),\pi^{A}(2),...,\pi^{A}(n))\\
P_{B}^{1}  &  =\left(  \left\vert \mu_{\pi^{B}(1)}\right\rangle ,\left\vert
\mu_{\pi^{B}(2)}\right\rangle ,...,\left\vert \mu_{\pi^{B}(n)}\right\rangle
\right)  \equiv(\pi^{B}(1),\pi^{B}(2),...,\pi^{B}(n))
\end{align*}

Let us consider the unitary transformation $T$ which takes the state
$\left\vert \mu_{k}\right\rangle $ into the state $\left\vert \nu
_{k}\right\rangle $. This can be written in the form%

\[
T=\sum_{j=1}^{n}\left\vert \nu_{j}\right\rangle \left\langle \mu
_{j}\right\vert
\]

and we suppose that the preference relations for game 2 are given by this
transformation applied to the preferences of game 1 so that%

\begin{align*}
P_{A}^{2}  &  =\left(  T\left\vert \mu_{\pi^{A}(1)}\right\rangle ,T\left\vert
\mu_{\pi^{A}(2)}\right\rangle ,...,T\left\vert \mu_{\pi^{A}(n)}\right\rangle
\right)  \equiv(\pi^{A}(1),\pi^{A}(2),...,\pi^{A}(n))\\
P_{B}^{2}  &  =\left(  T\left\vert \mu_{\pi^{B}(1)}\right\rangle ,T\left\vert
\mu_{\pi^{B}(2)}\right\rangle ,...,T\left\vert \mu_{\pi^{B}(n)}\right\rangle
\right)  \equiv(\pi^{B}(1),\pi^{B}(2),...,\pi^{B}(n))
\end{align*}

\begin{figure}
\centerline{\includegraphics[bb=1in .5in 9in 5in,scale=0.5]{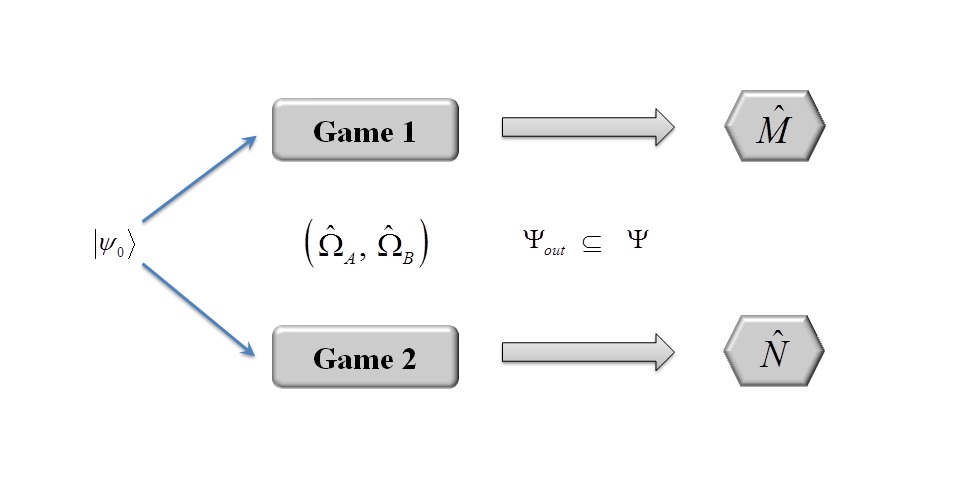}}
\caption{Two quantum games
characterized by the same input state and operations but different
measurements on the output state.}
\label{PGWQM_Fig3_3}
\end{figure}

It is clear that the players need to choose different strategies for each
game, in general. This can be most clearly seen if we assume that $\hat{M}%
$\ and $\hat{N}$\ are maximally conjugate. Let us further assume that the
operator sets for the players are such that they produce eigenstates of
$\hat{M}$\ in game 1. In game 2, these same operations lead to states that
have equal probability amplitudes, up to a phase factor, in the eigenstates of
$\hat{N}$. Thus in game 2 with these sets of operators there is no preferred
strategy since all strategies lead to the same expected outcome (this outcome
is $(n+1)/2$ if we adopt the weighting convention for the outcomes of the
previous sections).

It is an interesting question as to whether the uncertainty principle itself
can be cast in the form of a game. In this viewpoint we would treat the
different measurements as competing in some sense. Of course we would have to
consider the more general form of a quantum game in which the players were
allowed to perform a measurement. As we have seen in the above example,
localizing the output state to a particular eigenstate in game 1 produces a
corresponding uncertainty in the measurement of game 2 for that output state.
This will lead to a corresponding spread in the outcomes for game 2 when the
output state is localized on an eigenstate of game 1. Intuitively we would
expect that as we reduced the variance in the outcomes for game 1 we would
increase those of game 2 and vice versa.

Let us suppose that the output state for game 1 is the state $\left\vert
\psi\right\rangle $ the expected outcomes for player \textit{A} in games 1 and
2 with this output state are%

\begin{align*}
E_{1}^{A}\left(  \left\vert \psi\right\rangle \right)   &  =\sum_{k=1}%
^{p}\varpi_{\pi^{A}(k)}\left\vert \left\langle \mu_{k}|\psi\right\rangle
\right\vert ^{2}\\
E_{2}^{A}\left(  \left\vert \psi\right\rangle \right)   &  =\sum_{k=1}%
^{p}\varpi_{\pi^{A}(k)}\left\vert \left\langle \nu_{k}|\psi\right\rangle
\right\vert ^{2}%
\end{align*}

and we have labelled the numerical weights attached to the outcomes as
$\varpi_{j}$ with $\varpi_{1}>\varpi_{2}>...>\varpi_{p}$. If we define the
Hermitian operators $\hat{E}_{1}$ and $\hat{E}_{2}$ by%

\begin{align*}
\hat{E}_{1}  &  =f\left(  \hat{M}\right)  =\sum_{j=1}^{p}\varpi_{\pi^{A}%
(j)}\left\vert \mu_{j}\right\rangle \left\langle \mu_{j}\right\vert \\
\hat{E}_{2}  &  =f\left(  \hat{N}\right)  =\sum_{j=1}^{p}\varpi_{\pi^{A}%
(j)}\left\vert \nu_{j}\right\rangle \left\langle \nu_{j}\right\vert
\end{align*}

then the product of the variances of the outcomes in the two games satisfies
the uncertainty relation%

\[
\left(  \Delta\hat{E}_{1}\right)  ^{2}\left(  \Delta\hat{E}_{2}\right)
^{2}\geq\frac{1}{4}\left\vert \left\langle \psi\right\vert \left[  f\left(
\hat{M}\right)  ,f\left(  \hat{N}\right)  \right]  \left\vert \psi
\right\rangle \right\vert ^{2}%
\]

Of course the function $f$ is defined by a look-up table being a function that
has eigenvalues that are the weights arranged according to the preference relations.

\subsection{Quantum Dynamics as a Game}

The solution to the time-independent Schr\"{o}dinger equation can be expressed by%

\[
\left\vert \psi(t)\right\rangle =e^{-i\hat{H}t}\left\vert \psi
(0)\right\rangle
\]

where we have set $\hbar=1$. The time-evolution operator $\hat{T}=e^{-i\hat
{H}t}$\ is unitary. It is clear that this operator can be factored into the
product of two unitary operators $\hat{\alpha}$ and $\hat{\beta}$ so that
$\hat{T}=\hat{\beta}\hat{\alpha}$. There is obviously no unique way to perform
this factorisation. The evolution of the initial state $\left\vert
\psi(0)\right\rangle $ is fixed by the Hamiltonian and yields the output state
$\left\vert \psi(t)\right\rangle $. This output state can be thought of as the
equilibrium state of \textit{some} quantum game in which $\hat{\alpha}$ and
$\hat{\beta}$ represent the optimum strategy choices for the two players. Thus
we can think of the time evolution of a quantum system as being the result of
some game between two players. Constructing such a game and the associated
preference relations and sets $\hat{\Omega}_{A}$ and $\hat{\Omega}_{B}$ is
not, however, a trivial task. There are, in general, an infinite number of
such games for any quantum evolution that takes $\left\vert \psi
(0)\right\rangle $ to $\left\vert \psi(t)\right\rangle $.

It is clear from our previous discussion that if either of the players has
access to the full set of quantum mechanically allowed operations on a given
physical system used in a playable game then an equilibrium state cannot be
reached. If we are to associate a playable game with the evolution of some
state, then the operations available to the players are restricted by the form
of the Hamiltonian. In other words the Hamiltonian restricts the set of output
states. Note that whilst we can always associate some game with the time
evolution of a physical system, it may not always be possible to construct a
physically meaningful Hamiltonian that will yield the equilibrium state of a
given two-player game. However, if we can do this, then the solution of the
time-independent Schr\"{o}dinger equation yields the equilibrium state for
that game.

\section{2-Player 2-Qubit Games}

In what follows we shall consider a simple quantum system which can be used to
play a variety of playable quantum games. In this system we imagine that two
spin-1/2 particles are prepared in some state and input into some device which
allows the players to each perform a unitary operation on these particles (or
a sequence of such operations). The spin-1/2 particles are then transmitted to
some measurement device. The result of this measurement determines the
outcomes for the players.

We shall label the particles with \textit{A} and \textit{B} although this is
not to be understood that particle \textit{A} is only operated upon by player
\textit{A} necessarily, and similarly for particle \textit{B}. In the
spin-\textit{z} basis where the spin-up and spin-down states are labelled with
1 and 0, respectively, the general state of the two particles can be written as%

\[
\left\vert \psi\right\rangle =\alpha\left\vert 0,0\right\rangle +\beta
\left\vert 0,1\right\rangle +\gamma\left\vert 1,0\right\rangle +\delta
\left\vert 1,1\right\rangle
\]

where we have written, for example, $\left\vert 0,0\right\rangle =\left\vert
0\right\rangle _{z,A}\otimes\left\vert 0\right\rangle _{z,B}$. We shall drop
the labels \textit{z}, \textit{A} and \textit{B} for convenience, except where
ambiguity might arise. The particular quantum game that is played with this
physical system is determined by the input state of the two spin-1/2
particles, $\left\vert \psi_{0}\right\rangle $, the set of operations
$\hat{\Omega}_{A(B)}$ available to player \textit{A}(\textit{B}), the
Hermitian operator characterizing the measurement, $\hat{M}$, and the
preference relations of player \textit{A}(\textit{B}), $P_{A(B)}$, over the
eigenstates of the measurement operator $\hat{M}$.

\subsection{Playing a Classical Game with Quantum Coins}

Let us suppose that the initial state of our particles is given by $\left\vert
\psi_{0}\right\rangle =\left\vert 0,0\right\rangle $ and that the measurement
is simply the independent determination of the spin in the \textit{z}%
-direction of both particles so that \ $\hat{M}=\hat{\sigma}_{z}^{A}%
\otimes\hat{\sigma}_{z}^{B}$. The possible results of the measurement are thus
given by the eigenstates of $\hat{M}$ which are given by%

\begin{align*}
\left\vert \varphi_{1}\right\rangle  &  =\left\vert 0,0\right\rangle \\
\left\vert \varphi_{2}\right\rangle  &  =\left\vert 0,1\right\rangle \\
\left\vert \varphi_{3}\right\rangle  &  =\left\vert 1,0\right\rangle \\
\left\vert \varphi_{4}\right\rangle  &  =\left\vert 1,1\right\rangle
\end{align*}

Each player will have some (different) preference relation over these states.
We now suppose that the device that implements the game will only allow the
players to flip the spin of one particle and we further suppose that player
\textit{A}(\textit{B}) is restricted to operations only on particle
\textit{A}(\textit{B}). One possible representation of the sets of operations
available to the players is therefore given by%

\begin{align*}
\hat{\Omega}_{A}  &  =\{\hat{I},\exp\left(  i\frac{\pi}{2}\hat{\sigma}_{x}%
^{A}\right)  \}\\
\hat{\Omega}_{B}  &  =\{\hat{I},\exp\left(  i\frac{\pi}{2}\hat{\sigma}_{x}%
^{B}\right)  \}
\end{align*}

where $\hat{I}$ is the identity operator and $\hat{\sigma}_{x}^{A(B)}$ is the
spin operator in the \textit{x}-direction for particle \textit{A}(\textit{B}).

The preference relations for the players can be written as an ordered list of
the numbers 1,2,3, and 4 in an obvious shorthand notation. For example, the
preference relations%

\begin{align*}
P_{A}  &  =(2,1,4,3)\\
P_{B}  &  =(3,1,4,2)
\end{align*}

are those for the iconic game of Prisoner's Dilemma. The preference relations
alone do not completely specify a game, however, and the strategy set for the
players must also be taken into account. In classical games where there are
two choices in the strategy sets $S^{A(B)}$ so that there are 4 possible
outcomes, there are 432 possible pairs of preference relations in which the
first members of the lists differs. The quantum system we have described, with
the strategy set limited to a spin flip or identity operation, can implement
all 432 of these classical games as there is a direct one-to-one
correspondence between the various elements of the classical and quantum
games, that is%

\begin{align*}
\text{the result }\left\vert 0,0\right\rangle  &  \longleftrightarrow\text{
outcome }(0,0)\text{ \textit{etc}}\\
\hat{I}  &  \longleftrightarrow\text{ the choice 0}\\
\exp\left(  i\frac{\pi}{2}\hat{\sigma}_{x}^{A(B)}\right)   &
\longleftrightarrow\text{the choice 1}%
\end{align*}

The classical games can be implemented by giving each player a coin prepared
in a known state. The quantum version, with its restricted set of operators
for the players, uses `quantum coins' and can be thought of as simply an
`expensive' implementation of the corresponding game played with classical
coins. Even though there is a direct correspondence, and in fact the players
would just be playing the same game albeit with more complicated objects, it
is still correct to describe the quantum implementation as a quantum game
because it utilises quantum mechanical objects, operations and measurements.
In other words the physical implementation is the primary determinant of
whether a game is described as classical or quantum. Of course in this case
where we have restricted the available operations to flip or don't flip, the
results obtained by playing the quantum game are identical with those obtained
by playing the classical implementation. It should be obvious that the games
played with just two classical coins form a subset of the possible games that
can be played with the quantum implementation when we ease the restriction on
the set of available operations.

It is tempting to describe this procedure as `quantizing a game' but this
terminology is fraught with difficulties. It would certainly be correct to say
we have quantized the physical system that is used to implement a game in
going from classical to quantum coins. However, a game is also characterized
by a set of available strategies and by enlarging the strategy set in a
quantum mechanical description, for example, we are, in effect, playing a
different game in classical terms. That being said, it would be legitimate to
describe a game played with quantum coins with the preference relations of
Prisoner's Dilemma over the 4 outcomes described by the chosen measurement
operator as a single quantum game with many different playable versions
according to how the available operations and input states are restricted. We
shall examine in what sense it is possible to talk of `quantizing a game'
after we have looked at some more games that can be played with our quantum implementation.

\subsection{Quantum Games and Mixed Strategies}

In the previous quantum game we only allowed the players operations that
resulted in the production of eigenstates of the measurement operator. In
general this will not be the case and the resultant output state from the
device that implements the players choices will not be an eigenstate of the
measurement operator. The action of measurement will therefore, in general,
lead to some distribution over the eigenstates resulting in a distribution
over the outcomes for the players.

Of course, distributions over the parameters of a classical game are an
integral part of game theory. In terms of our \textit{playable} classical game
there are at least 3 ways that a distribution over the outcomes can emerge. We
could arrange our classical system such that there is some distribution over
the initial state of the game. If the game were to be implemented using coins,
for example, then there would be some distribution over the initial state of
the coins and the players would select their subsequent strategy accordingly.
Alternatively, the initial state of the system could be determined, but the
players could select their strategies according to some distribution. This
kind of game is usually termed a \textit{mixed} game, or one in which the
players adopt a mixed strategy. A game in which the players select their
strategies deterministically is usually termed a \textit{pure} game, or one in
which the players adopt pure strategies. The third possibility is that the
outcomes for the players are determined according to some distribution. Each
of these possibilities will lead to a distribution over the outcomes for the
players and they must consider their \textit{expected} outcome from the game
when determining their choice of strategy or distribution over those strategies.

In a playable quantum game each of these elements could also be present.
However, quantum mechanics, through measurement, forces a distribution over
the outcomes, in general, even when there is no other element of stochasticity
introduced in the game. A distribution over the outcomes will occur in a
quantum game whenever the output state from the players is not an eigenstate
of the measurement operator. In general the output state of the players will
be of the form $\left\vert \psi\right\rangle =\alpha\left\vert
0,0\right\rangle +\beta\left\vert 0,1\right\rangle +\gamma\left\vert
1,0\right\rangle +\delta\left\vert 1,1\right\rangle $, as above, where the
probability of obtaining a particular result upon measurement is given by the
square modulus of the amplitudes. Thus, in general, each possible output state
from the players will lead to a different distribution of outcomes in a
quantum game. There is an expected outcome for each possible output state.
This is quite different from the situation in a classical mixed game in which
the players select their strategies with some distribution. We could model the
quantum game with a classical game if, in the classical game, the outcomes
were computed by taking the measurement as input to some function that
generates a different distribution of outcomes for each possible output of the
players. In the quantum case the distribution over the outcomes for each
possible state arises from the physical properties of quantum measurement. In
the classical case these distributions arise as the result of a
\textit{computation}. So, provided that the function that generates the
different distributions in the classical case is computable, we can always
model a quantum game with some appropriate classical game. We shall consider a
simple example of this shortly.

There are at least 3 ways in which we can extend the simple quantum game of
the previous section so that the resultant output state is not an eigenstate
of the measurement operator. Let us briefly examine each of these possibilities.

\subsubsection{Rotation of the input state}

As in the previous section we consider a quantum game in which the measurement
is simply the independent determination of the spin in the \textit{z}%
-direction of both particles so that \ $\hat{M}=\hat{\sigma}_{z}^{A}%
\otimes\hat{\sigma}_{z}^{B}$. The device that implements the game will only
allow the players to flip the spin of one particle in the \textit{z}-direction
and we further suppose that player \textit{A}(\textit{B}) is restricted to
operations only on particle \textit{A}(\textit{B}). Now, however, we consider
the input states to have undergone some rotation before they are acted upon by
the players. The input state is now of the form $\left\vert \psi
_{0}\right\rangle =\left\vert 0,0\right\rangle _{\theta,\theta^{\prime
},\varphi,\varphi^{\prime}}=\left\vert 0\right\rangle _{\theta,\varphi
,A}\otimes\left\vert 0\right\rangle _{\theta^{\prime},\varphi^{\prime},B}$
where the rotated state of particle A is given by%

\[
\left\vert 0\right\rangle _{\theta,\varphi,A}=\cos\left(  \theta/2\right)
e^{i\varphi/2}\left\vert 0\right\rangle _{z}-\sin\left(  \theta/2\right)
e^{-i\varphi/2}\left\vert 1\right\rangle _{z}%
\]

with a similar expression for particle B. The input state is an eigenstate of
the spin operator%

\[
\hat{\sigma}\left(  \theta,\varphi\right)  =\cos\theta~\hat{\sigma}_{z}%
+\sin\theta~e^{-i\varphi}\hat{\sigma}_{+}+\sin\theta~e^{i\varphi}\hat{\sigma
}_{-}%
\]

where we have written the spin raising and lowering operators in the
\textit{z}-direction as $\hat{\sigma}_{\pm}$. If we write $c=\cos\left(
\theta/2\right)  e^{i\varphi/2}$ and $s=\sin\left(  \theta/2\right)
e^{-i\varphi/2}$, then, in terms of the eigenstates $\left\vert \varphi
_{k}\right\rangle $ of $\hat{M}$, the four possible output states are given by%

\begin{align*}
\left\vert \psi_{11}\right\rangle  &  =cc^{\prime}\left\vert \varphi
_{1}\right\rangle -cs^{\prime}\left\vert \varphi_{2}\right\rangle -sc^{\prime
}\left\vert \varphi_{3}\right\rangle +ss^{\prime}\left\vert \varphi
_{4}\right\rangle \\
\left\vert \psi_{12}\right\rangle  &  =-cs^{\prime}\left\vert \varphi
_{1}\right\rangle +cc^{\prime}\left\vert \varphi_{2}\right\rangle +ss^{\prime
}\left\vert \varphi_{3}\right\rangle -sc^{\prime}\left\vert \varphi
_{4}\right\rangle \\
\left\vert \psi_{21}\right\rangle  &  =-sc^{\prime}\left\vert \varphi
_{1}\right\rangle +ss^{\prime}\left\vert \varphi_{2}\right\rangle +cc^{\prime
}\left\vert \varphi_{3}\right\rangle -cs^{\prime}\left\vert \varphi
_{4}\right\rangle \\
\left\vert \psi_{22}\right\rangle  &  =ss^{\prime}\left\vert \varphi
_{1}\right\rangle -sc^{\prime}\left\vert \varphi_{2}\right\rangle -cs^{\prime
}\left\vert \varphi_{3}\right\rangle +cc^{\prime}\left\vert \varphi
_{4}\right\rangle
\end{align*}

and we have taken the indices for the output state such that 1 is the identity
and 2 is the spin-flip operation so that $\left\vert \psi_{21}\right\rangle $
is the output state when player \textit{A} chooses to flip and player
\textit{B} does not, for example. Note that because of the special structure
of this game in which the operations available to the players merely transform
between eigenstates of the measurement operator, there are only 4 distinct
amplitudes that occur in these output states.

Let us label the probabilities $p_{i}$ as follows : $p_{1}=\left\vert
cc^{\prime}\right\vert ^{2},p_{2}=\left\vert cs^{\prime}\right\vert ^{2}%
,p_{3}=\left\vert sc^{\prime}\right\vert ^{2}$, and $p_{4}=\left\vert
ss^{\prime}\right\vert ^{2}$. The expected outcomes for player \textit{A} with
the preference relations of Prisoner's Dilemma, $P_{A}=(2,1,4,3)$, are given by%

\begin{align*}
E_{11}^{A}  &  =4\left\vert cs^{\prime}\right\vert ^{2}+3\left\vert
cc^{\prime}\right\vert ^{2}+2\left\vert ss^{\prime}\right\vert ^{2}+\left\vert
sc^{\prime}\right\vert ^{2}=4p_{2}+3p_{1}+2p_{4}+p_{3}\\
E_{12}^{A}  &  =4\left\vert cc^{\prime}\right\vert ^{2}+3\left\vert
cs^{\prime}\right\vert ^{2}+2\left\vert sc^{\prime}\right\vert ^{2}+\left\vert
ss^{\prime}\right\vert ^{2}=4p_{1}+3p_{2}+2p_{3}+p_{4}\\
E_{21}^{A}  &  =4\left\vert ss^{\prime}\right\vert ^{2}+3\left\vert
sc^{\prime}\right\vert ^{2}+2\left\vert cs^{\prime}\right\vert ^{2}+\left\vert
cc^{\prime}\right\vert ^{2}=4p_{4}+3p_{3}+2p_{2}+p_{1}\\
E_{22}^{A}  &  =4\left\vert sc^{\prime}\right\vert ^{2}+3\left\vert
ss^{\prime}\right\vert ^{2}+2\left\vert cc^{\prime}\right\vert ^{2}+\left\vert
cs^{\prime}\right\vert ^{2}=4p_{3}+3p_{4}+2p_{1}+p_{2}%
\end{align*}

Player \textit{B} has the preference relation $P_{B}=(3,1,4,2)$ for Prisoner's
dilemma. The expected outcomes for player \textit{B} for each of the possible
ouput states are therefore given by%

\begin{align*}
E_{11}^{B}  &  =4\left\vert sc^{\prime}\right\vert ^{2}+3\left\vert
cc^{\prime}\right\vert ^{2}+2\left\vert ss^{\prime}\right\vert ^{2}+\left\vert
cs^{\prime}\right\vert ^{2}=4p_{3}+3p_{1}+2p_{4}+p_{2}\\
E_{12}^{B}  &  =4\left\vert ss^{\prime}\right\vert ^{2}+3\left\vert
cs^{\prime}\right\vert ^{2}+2\left\vert sc^{\prime}\right\vert ^{2}+\left\vert
cc^{\prime}\right\vert ^{2}=4p_{4}+3p_{2}+2p_{3}+p_{1}\\
E_{21}^{B}  &  =4\left\vert cc^{\prime}\right\vert ^{2}+3\left\vert
sc^{\prime}\right\vert ^{2}+2\left\vert cs^{\prime}\right\vert ^{2}+\left\vert
ss^{\prime}\right\vert ^{2}=4p_{1}+3p_{3}+2p_{2}+p_{4}\\
E_{22}^{B}  &  =4\left\vert cs^{\prime}\right\vert ^{2}+3\left\vert
ss^{\prime}\right\vert ^{2}+2\left\vert cc^{\prime}\right\vert ^{2}+\left\vert
sc^{\prime}\right\vert ^{2}=4p_{2}+3p_{4}+2p_{1}+p_{3}%
\end{align*}

This game can be implemented entirely classically as follows. We give each of
the players a coin prepared in the state T. The players are allowed to not
flip, \textit{\={F}}, or flip, \textit{F}, their respective coins. When the
measurement device receives the two coins in the state (H,T), for example, it
will \textit{compute} an outcome for player \textit{A} of 4 with probability
$p_{4}$, an outcome of 3 with probability $p_{3}$, an outcome of 2 with
probability $p_{2}$ and an outcome of 1 with probability $p_{1}$. The outcomes
for player \textit{B} are computed to be 4 with probability $p_{1}$, an
outcome of 3 with probability $p_{3}$, an outcome of 2 with probability
$p_{2}$ and an outcome of 1 with probability $p_{4}$. In the classical
version, therefore, the probabilities induced by measurement in the quantum
game are replaced by a computation. If the probabilities induced by the
measurement in the quantum game are computable then any quantum game in which
the players have a finite set of strategies can be replaced by an equivalent
appropriate classical game played with coins in which the outcomes are
determined by assignment of the computed probabilities.

\subsubsection{Rotation of the operators}

We now consider that the input state remains unchanged, being the product of
the spin-down states of the two particles in the \textit{z}-direction as in
section 3.1. The measurement operator remains unchanged and is $\hat{M}%
=\hat{\sigma}_{z}^{A}\otimes\hat{\sigma}_{z}^{B}$ as before. However, we now
rotate the available operations of the players so that their spin flip is
performed along a different axis. This spin flip operation will have the
effect of rotating the input state of the particles so that the output state
will no longer be an eigenstate of the measurement. The flip operation for
player \textit{A} is along the direction defined by the spin operator
$\hat{\sigma}\left(  \theta,\varphi\right)  $, and that of player \textit{B}
along the direction defined by $\hat{\sigma}\left(  \theta^{\prime}%
,\varphi^{\prime}\right)  $.

It is easy to see that this generates different probabilities than when we
rotate the input state. Consider the choice where both players do not flip.
The resultant output state $\left\vert \psi_{11}\right\rangle $\ is simply the
original input state and this is an eigenstate of the measurement operator
leading to a particular outcome with probability 1. The other possible output
states are superpositions dependent upon the angle of rotation. In general,
the probabilities occuring in the determination of the outcomes are functions
of the angles $\theta,\theta^{\prime},\varphi,\varphi^{\prime}$. The overall
quantum game with this set of operators for the players can, again, be
modelled by an entirely classical game played with just two coins for the players.

\subsubsection{Rotation of the measurement}

In this case the input state remains as before being $\left\vert \psi
_{0}\right\rangle =\left\vert 0\right\rangle _{z,A}\otimes\left\vert
0\right\rangle _{z,B}$, the spin-flip allowed to the players is in the
\textit{z}-direction, but the measurement is rotated so that it becomes
$\hat{M}_{\theta,\theta^{\prime},\varphi,\varphi^{\prime}}=\hat{\sigma}%
^{A}\left(  \theta,\varphi\right)  \otimes\hat{\sigma}^{B}\left(
\theta^{\prime},\varphi^{\prime}\right)  $. It is clear that this is entirely
physically equivalent to the game played with an input state rotated through
the angles $\theta,\theta^{\prime},\varphi,\varphi^{\prime}$ considered in
section 3.2.1. The same probabilities for the outcomes will be obtained.

\subsubsection{An entangled input state}

A potentially more interesting case occurs when the input state is entangled.
Such states cannot be represented by any collection of classical objects, such
as coins. Let us look at the question of whether, nevertheless, the resulting
quantum game can still be simulated with classical coins. The most
`non-classical' states of 2 spin-1/2 particles are the maximally entangled
states. These states give maximal violations of Bell's inequality. Let us
consider an input state of the form $\left\vert \psi_{0}\right\rangle
=\frac{1}{\sqrt{2}}\left(  \left\vert 0\right\rangle _{z,A}\otimes\left\vert
1\right\rangle _{z,B}+\left\vert 1\right\rangle _{z,A}\otimes\left\vert
0\right\rangle _{z,B}\right)  $ with the players allowed to spin-flip in the
\textit{z}-direction and the measurement given by $\hat{M}=\hat{\sigma}%
_{z}^{A}\otimes\hat{\sigma}_{z}^{B}$. The 4 maximally-entangled Bell states
that form a basis for the particles are, of course, eigenstates of $\hat{M}$.
The 4 possible output states the players can produce are given by%

\begin{align*}
\left\vert \psi_{11}\right\rangle  &  =\frac{1}{\sqrt{2}}\left(  \left\vert
0,1\right\rangle +\left\vert 1,0\right\rangle \right)  =\frac{1}{\sqrt{2}%
}\left(  \left\vert \varphi_{2}\right\rangle +\left\vert \varphi
_{3}\right\rangle \right) \\
\left\vert \psi_{12}\right\rangle  &  =\frac{1}{\sqrt{2}}\left(  \left\vert
0,0\right\rangle +\left\vert 1,1\right\rangle \right)  =\frac{1}{\sqrt{2}%
}\left(  \left\vert \varphi_{1}\right\rangle +\left\vert \varphi
_{4}\right\rangle \right) \\
\left\vert \psi_{21}\right\rangle  &  =\frac{1}{\sqrt{2}}\left(  \left\vert
1,1\right\rangle +\left\vert 0,0\right\rangle \right)  =\frac{1}{\sqrt{2}%
}\left(  \left\vert \varphi_{1}\right\rangle +\left\vert \varphi
_{4}\right\rangle \right) \\
\left\vert \psi_{22}\right\rangle  &  =\frac{1}{\sqrt{2}}\left(  \left\vert
1,0\right\rangle +\left\vert 0,1\right\rangle \right)  =\frac{1}{\sqrt{2}%
}\left(  \left\vert \varphi_{2}\right\rangle +\left\vert \varphi
_{3}\right\rangle \right)
\end{align*}

We see that there are only 2 distinct output states that can be produced. In
other words, the measurement $\hat{M}=\hat{\sigma}_{z}^{A}\otimes\hat{\sigma
}_{z}^{B}$ is not sufficient to properly distinguish between the moves of the
players. This is because the measurement cannot distinguish between the 4 Bell
states, which give a degeneracy. If we only allow the players to act on their
`own' particle with a unitary transformation the output states will always be
maximally entangled. This is because independent unitary operations on the
particles cannot reduce the degree of entanglement. In order to properly
distinguish between the output states we need to perform a Bell-type
measurement in which joint properties of the output states are measured.

Let us, once more, consider the preference relations of Prisoner's dilemma so
that $P_{A}=(2,1,4,3)$ and $P_{B}=(3,1,4,2)$. We let the actual outcomes, that
is, the numerical value of the rewards be the numbers \textit{a}, \textit{b},
\textit{c} and \textit{d} where $a>b>c>d$. In terms of these numbers the
rewards for the players are given by%

\begin{align*}
\text{measurement result }\left\vert \varphi_{1}\right\rangle  &
\longrightarrow(b,b)\\
\text{measurement result }\left\vert \varphi_{2}\right\rangle  &
\longrightarrow(a,d)\\
\text{measurement result }\left\vert \varphi_{3}\right\rangle  &
\longrightarrow(d,a)\\
\text{measurement result }\left\vert \varphi_{4}\right\rangle  &
\longrightarrow(c,c)
\end{align*}

The expected outcomes for the players are therefore given by%

\begin{align*}
E_{11}^{A}  &  =\frac{1}{2}(a+d)=E_{11}^{B}\\
E_{12}^{A}  &  =\frac{1}{2}(b+c)=E_{12}^{B}\\
E_{21}^{A}  &  =\frac{1}{2}(b+c)=E_{21}^{B}\\
E_{22}^{A}  &  =\frac{1}{2}(a+d)=E_{22}^{B}%
\end{align*}

In some sense the game of Prisoner's Dilemma is a pathological example because
of the symmetries involved. If the rewards are such that $a+d=c+b$ (as is the
case with the assignment of numerical rewards we have previously used) then
the expected outcomes are all equal and thus there is no rational reason for
the players to select one move above another. If, on the other hand, the
rewards are such that $a+d>c+b$ then rational players would clearly prefer
either of the output states $\left\vert \psi_{11}\right\rangle $ or
$\left\vert \psi_{22}\right\rangle $. There is no way of ensuring this in the
absence of communication between the players. Similar remarks apply if we have
the condition $a+d<c+b$ but now the players would prefer either of the output
states $\left\vert \psi_{12}\right\rangle $ or $\left\vert \psi_{21}%
\right\rangle $. We see exactly the same inability to select an advantageous
strategy if we play the game of quantum chicken with the same entangled input
state and the same strategy sets for the players.

The closest classical input to the entangled quantum state would be to arrange
the game so that player \textit{A} would receive a coin in the state H or T
uniformly at random but with the coin for player \textit{B} being prepared in
the opposite state. This does not reproduce the same expected outcomes, but
does lead to a situation in which the selection of a strategy cannot improve
the results for either player, just as in the quantum case. The expected
outcomes cannot be reproduced in a\ mixed classical game either. It is only by
adapting the function that determines the outcomes according to the
probabilities from the quantum version that we can simulate this game with
classical coins. The function for simulating QPD$_{\hat{\sigma}_{z}^{A}%
\otimes\hat{\sigma}_{z}^{B}}$ with the above strategy sets and entangled input
can be represented by the following rules%

\begin{align*}
\text{output state (T,T) }  &  \text{:}p_{A}(a)=p_{A}(d)=\frac{1}{2}%
=p_{B}(a)=p_{B}(d)\\
\text{output state (T,H) }  &  \text{:}p_{A}(b)=p_{A}(c)=\frac{1}{2}%
=p_{B}(b)=p_{B}(c)\\
\text{output state (H,T) }  &  \text{:}p_{A}(b)=p_{A}(c)=\frac{1}{2}%
=p_{B}(b)=p_{B}(c)\\
\text{output state (H,H) }  &  \text{:}p_{A}(a)=p_{A}(d)=\frac{1}{2}%
=p_{B}(a)=p_{B}(d)
\end{align*}

where the input state to the game is (T,T) and $p_{A}(a)$\ is the probability
that player \textit{A} receives the reward $a$, and so on.

\subsection{Quantizing a Game?}

As we have seen, a 2-player game is described by a set of objects, commonly
called strategies, from which a tuple is chosen. These tuples become the input
to some function that determines the `rewards' for the players. Armed with
knowledge of this function the players select their element of the strategy
tuple in such a way as to attempt to optimise their reward, taking into
account the fact that the other player is making exactly the same
deliberation. Physics deals with objects that can be measured and manipulated
in the `real' world. Accordingly, if physics has anything at all to say about
game theory, and vice versa, we must consider what it actually means to play a
game in terms of actions and measurements on real physical objects. In other
words we require our games to be physically realizable, that is,
\textit{playable}, if we are to elucidate any potential impact that the
underlying physics might have on game theory. Alternatively we might also look
for new insights into a physical theory by application of the techniques of
game theory. Again, however, we must look to games that are physically
realizable if we are to determine whether game theory can shed any new light
on a physical theory.

In the quantum theory of light it is an important consideration to determine
whether any predicted effects are truly quantum in nature. That is, we try to
determine whether the predicted behaviour occurs only as a result of the
quantization of the electromagnetic field, or whether treating the field
classically is sufficient to explain any prediction. It would be a natural
question, therefore, to ask whether there are any `non-classical' effects that
can be obtained by playing a game with quantum objects. If the properties of a
quantum game can be described by \textit{some} game implemented entirely with
classical objects then in what sense could the quantum version be said to be
truly dependent on quantum effects? We would argue that the answer to this
question is \textquotedblleft\textit{not at all}\textquotedblright. The
classical version may not look much like the quantum game, but if
\textit{some} equivalent classical implementation can be found then it seems
clear to us that there are no genuinely quantum-mechanical effects occurring
in the quantum game [16].

Let us consider a general 2-player quantum game in which the input state is
$\left\vert \psi_{0}\right\rangle $\ and the players select an operation from
finite sets $\hat{\Omega}_{A}$\ and $\hat{\Omega}_{B}$ with $\left\vert
\hat{\Omega}_{A}\right\vert =n_{A}$\ and $\left\vert \hat{\Omega}%
_{B}\right\vert =n_{B}$. There are $n_{A}n_{B}$\ possible output states given by%

\[
\left\vert \psi_{ij}\right\rangle =\hat{\beta}_{j}\hat{\alpha}_{i}\left\vert
\psi_{0}\right\rangle =\sum_{k=1}^{p}\left\langle \varphi_{k}\left\vert
\hat{\beta}_{j}\hat{\alpha}_{i}\right\vert \psi_{0}\right\rangle \left\vert
\varphi_{k}\right\rangle
\]

where we have assumed that player \textit{A} moves first and that the
eigenstates of the measurement operator are given by $\left\vert \varphi
_{k}\right\rangle $. The expected outcomes for the players are given, as
before, as%

\begin{align*}
E_{ij}^{A}  &  =\sum_{k=1}^{p}(n-k+1)\left\vert \left\langle \varphi_{\pi
^{A}(k)}|\hat{\beta}_{j}\hat{\alpha}_{i}|\psi_{0}\right\rangle \right\vert
^{2}\\
E_{ij}^{B}  &  =\sum_{k=1}^{p}(n-k+1)\left\vert \left\langle \varphi_{\pi
^{B}(k)}|\hat{\beta}_{j}\hat{\alpha}_{i}|\psi_{0}\right\rangle \right\vert
^{2}%
\end{align*}

The players use their calculation of the expected outcomes to determine their
selection of appropriate operation. The probabilities $P_{ij}^{A(B)}%
=\left\vert \left\langle \varphi_{\pi^{A(B)}(k)}|\hat{\beta}_{j}\hat{\alpha
}_{i}|\psi_{0}\right\rangle \right\vert ^{2}$ can be calculated by the players
if they have full knowledge of the game parameters.

Now let us consider a classical game implemented entirely by classical coins.
Player \textit{A}(\textit{B}) is given $\kappa_{A(B)}$ coins where for
convenience we shall take $n_{A(B)}=2^{\kappa_{A(B)}}$.The players can flip
any number of coins or choose not to flip. In effect the players are selecting
a codeword that represents the choice of an element from a set of cardinality
$n_{A(B)}$. Recall that the outcomes for the players in a classical game are
determined by the computation of some function that assigns rewards according
to the input tuple that represents the selection of the players' strategies.
Thus if this function is chosen so that the rewards are assigned probabilities
according to those determined by the calculated probabilities of a quantum
game we can use the classical game to simulate the quantum version.

It is clear that, provided the probabilities of a given quantum game can be
\textit{calculated}, we can simulate that quantum game by some classical game
played entirely with classical coins. Thus, in the sense we have argued for
above, there are no results obtainable by playing a game using
quantum-mechanical objects that cannot be simulated by playing \textit{some}
game with classical coins. In this sense, therefore, we do not expect to see
any genuine non-classical results emerging from the process of considering
games played with quantum objects, as far as the calculation of the strategies
and the determination of the expected outcomes considered here. It should be
noted, however, that if the measurement is more sophisticated and the outcomes
the players receive are calculated by considering the correlations in an
ensemble of games then we would not expect to always be able to simulate this
with classical coins when the quantum systems used to play the games violate a
Bell inequality [9-11].

There is an assumption here that the computation performed after the
measurement can be achieved classically. In order to see quantum effects, that
computation would have to depend intrinsically on quantum processes. In other
words, the computation that determined the output would itself have to be
quantum-mechanical in nature, and not reproducible by any classical computing
device, in order for us to see any genuine non-classical effect in a quantum
game in the sense we have described above.

With these remarks in mind we return to the question of what it means to
`quantize' a game. From the perspective of physics quantization really only
makes sense when we consider objects and entities that have some physical
existence and that existence is characterised by physical parameters. The
classical variables are replaced by operators and the state of the system is
described by a wavefunction, or more abstractly a vector in Hilbert space.
Thus it really only makes physical sense to talk of quantizing a game when
that game is described by physical objects. It is legitimate to talk of
quantizing the physical system that is used to implement a game, but is this
strictly equivalent to quantizing a game?

In mathematical terms a game can be abstractly described as a mapping of some
tuple onto a tuple of outcomes (one for each player). The domain of the
function that performs this mapping is the product of the strategy sets. Game
theory is all about what the players would choose from these strategy sets
given their knowledge of the potential outcomes and the potential actions of
the other player. If we extend the domain by allowing an enlarged set of moves
for the players (and, of course, extend the function to accomodate the
enlarged domain) then we are not really playing the same game in classical
terms, even if we we keep the same range for the mapping. The original,
smaller, game may be in some sense contained within this larger game, or we
might consider the enlarged game to be an extension of the original game, but
in neither case could we consider that we are playing the same game. So, if we
`quantize' a game and allow an extended set of operations from those of the
classical implementation we have quantized we are not really comparing like
with like.

One way to approach the quantization of a game is to consider a classical game
and to extend this into the quantum domain. In this way we guarantee that, if
the extension is correctly performed, the classical game sits within the
quantum game and can be obtained in the appropriate classical limit. The
advantage of this is that it maintains the link to the classical game in a
clear and compelling way [7]. This aproach is particularly useful if we are
interested in the question of how quantum mechanics may change the properties
of a particular classical game, if at all. \ However, we must be careful in
our comparison. As discussed above, if, in performing the extension, we allow
the players a greater range of `moves', we are not really playing an
equivalent game even if the classical game is properly contained within the
extension in the appropriate limit.

An alternative viewpoint, and the one we believe to be more fruitful, is to
consider the properties of playing a game with quantum objects, whether or not
any classical game of interest is properly contained within it in the
appropriate limit. We could describe this viewpoint as `gaming the quantum'. A
quantum game can be considered to be complete, with respect to a particular
measurement, if we consider the domain of the function that performs the
mapping to our output states to be the set of all possible states of the
physical system, together with the set of all possible unitary operations
allowed by quantum mechanics on those states. The range of this function is
clearly the set of all possible states of the physical system. A particular
game, with respect to a measurement, is then specified by the preference
relations on the eigenstates of the measurement operator.

It is important to note that a quantum game is specified with respect to a
measurement on a particular physical system. If we consider a 4-state quantum
system, for example, then there is a critical difference in whether these 4
states arise from a single quantum system, or whether they arise from the
product space of two 2-state systems. In the latter case we need to consider
entanglement between the two physical systems, whereas in the former there is
no issue of entanglement.

If we consider a 4-state quantum system which arises from the joint space of
two 2-state systems, and we suppose the preference relations of Prisoner's
Dilemma then the complete quantum game with respect to some measurement
contains all possible quantum variants of Prisoner's Dilemma that can be
played with that system. Accordingly, we would describe such a game as Quantum
Prisoner's Dilemma (QPD$_{\hat{M}}$) with respect to the measurement $\hat{M}%
$. If we restrict the domain of the system by restricting the available
operations of the players then we can play an infinite number of variants of
QPD$_{\hat{M}}$.

Not all of these variants will contain the classical game of Prisoner's
Dilemma in the appropriate limit. All of these variants, however, are
contained within the \textit{single} quantum game of QPD$_{\hat{M}}$. The
single complete quantum game QPD$_{\hat{M}}$, therefore, describes a family of
games characterized by the measurement and the preferences over the outcomes
of the measurement. The different members of this family can be generated by
suitable restrictions of the available operations of the players. In the next
section we examine some of the features of playing the quantum game
QPD$_{\hat{M}}$.

\section{Playing Quantum Prisoner's Dilemma}

The game of Quantum Prisoner's Dilemma with respect to a measurement $\hat{M}
$\ is played with two 2-state particles, conveniently represented as spin-1/2
particles. In the complete game we allow the players any unitary operation
consistent with the laws of quantum mechanics. Thus, the players are allowed
to perform any operation in the joint state space $H_{A}\otimes H_{B}$\ of the
two particles. As we have noted, the complete game with pure strategies cannot
produce an equilibrium state if the players have perfect information. If we
assume that there exists a preferred pure strategy for player \textit{A} then
this is determined by some computation of player \textit{A}. If player
\textit{A} can perform this computation then so can player \textit{B} and,
accordingly, player \textit{B} can simply apply the inverse operation which
contradicts our assumption of the existence of a preferred strategy.

In order for QPD$_{\hat{M}}$\ to produce some output state that is of interest
in game-theoretic terms, therefore, the possible output states must be
restricted. This means that the operator sets available to the players are
such that $\hat{\Omega}_{A(B)}\subset\hat{\Omega}$. The set of possible output
states can be partitioned into 3 disjoint sets as follows\medskip

$\Psi^{i}$ : the set of all output states that can be written as a tensor
product state of the two particles $\left\vert \psi_{A}\right\rangle
\otimes\left\vert \psi_{B}\right\rangle $

$\Psi_{\max}^{e}$ : the set of all output states that are maximally entangled

$\Psi_{p}^{e}$ : the set of all output states that have a partial degree of
entanglement\medskip

The set of all possible output states is simply the union of these 3 disjoint
sets $\Psi=\Psi^{i}\cup\Psi_{p}^{e}\cup\Psi_{\max}^{e}$. A necessary, but not
sufficient, condition for the quantum game to `contain' the classical game in
the appropriate limit is that the available operations must be capable of
producing output states $\left\vert \psi\right\rangle $ such that $\left\vert
\psi\right\rangle \in\Psi^{i}$. If we suppose that our players could choose to
play 4 different versions $G_{i}$ of our game such that\medskip

$G_{1}$ : produces output states $\left\vert \psi\right\rangle \in\Psi^{i} $

$G_{2}$ : produces output states $\left\vert \psi\right\rangle \in\Psi^{i}%
\cup\Psi_{p}^{e}$

$G_{3}$ : produces output states $\left\vert \psi\right\rangle \in\Psi_{p}%
^{e}$

$G_{4}$ : produces output states $\left\vert \psi\right\rangle \in\Psi_{\max
}^{e}\medskip$

then it is an interesting question as to whether rational players would choose
to play game 1,2,3, or 4. The games $G_{i}$ are produced by different
restrictions on the operator sets of the players but they are all instances of
the complete quantum game of QPD$_{\hat{M}}$. Of course, the partitioning of
the output space is not unique and there are other partitions we could
envisage. The one we have adopted, however, does allow us to place the
previous work on QPD$_{\hat{M}}$\ within the framework of playable quantum
games that we have developed so that we can see the connection between them
and how they arise naturally from the single quantum game of QPD$_{\hat{M}}%
$\ with the appropriate restrictions.

\subsection{Playing QPD$_{\hat{M}}$ with Maximally Entangled States}

We have already touched upon a version of QPD$_{\hat{M}}$\ when we input two
maximally entangled particles in section 3.2.4. In that case we only allowed
the players the two unitary operations of flip or don't flip in the
\textit{z}-direction.on one spin each. In the game envisioned by MW [8] the
players are allowed to rotate the state of their respective particles in an
arbitrary direction. As we have already noted, independent unitary
transformations cannot change the degree of entanglement and the ouput state
in this game is always maximally entangled. Thus the MW game is an instance of
a game of type $G_{4}$. Furthermore, the ability to generate any arbitrary
rotation ensures that the set of output states from this game is just
$\Psi_{\max}^{e}$, that is, $\Psi_{out}=\Psi_{\max}^{e}$.

In order to compare the various games we can play with QPD$_{\hat{M}}$\ we
shall keep the input state fixed as $\left\vert \psi_{0}\right\rangle
=\left\vert 0\right\rangle _{z,A}\otimes\left\vert 0\right\rangle _{z,B}$ and
the measurement as $\hat{M}=\hat{\sigma}_{z}^{A}\otimes\hat{\sigma}_{z}^{B}$.
By varying the operator sets available to the players we can restrict the
possible output states. Let us assume that player \textit{A} moves first. The
game with a maximally entangled state as input is therefore described by the
preference relations on the eigenstates of the measurement together with the
operator sets%

\begin{align*}
\left(  \hat{\Omega}_{A}T\right)   &  =\{\hat{\alpha}_{1}T,\hat{\alpha}%
_{2}T,...,\hat{\alpha}_{p}T\}\\
\hat{\Omega}_{B}  &  =\{\hat{\beta}_{1},\hat{\beta}_{2},...,\hat{\beta}_{q}\}
\end{align*}

where $T$ is an entanglement operator that generates the maximally entangled
state $\left\vert \xi\right\rangle =T\left\vert 0\right\rangle _{z,A}%
\otimes\left\vert 0\right\rangle _{z,B}=\frac{1}{\sqrt{2}}\left(  \left\vert
0\right\rangle _{z,A}\otimes\left\vert 1\right\rangle _{z,B}+\left\vert
1\right\rangle _{z,A}\otimes\left\vert 0\right\rangle _{z,B}\right)  $. If the
operator sets are restricted so that $\{\hat{\alpha}_{k}\}$ acts only on the
space $H_{A}$ and $\{\hat{\beta}_{k}\}$ acts only on the space $H_{B}$ then
the ouput states are always maximally entangled. The output states will also
be maximally entangled if we enlarge the sets $\{\hat{\alpha}_{k}\}$\ and
$\{\hat{\beta}_{k}\}$\ to include operations that can act on both spaces
independently, that is, without causing the particles to interact. This is a
game of the form MW$_{1A}$\ discussed in section 2.

If player \textit{B} makes the first move, then the game is of the form
MW$_{1B}$\ with the operator sets%

\begin{align*}
\hat{\Omega}_{A}  &  =\{\hat{\alpha}_{1},\hat{\alpha}_{2},...,\hat{\alpha}%
_{p}\}\\
\left(  \hat{\Omega}_{B}T\right)   &  =\{\hat{\beta}_{1}T,\hat{\beta}%
_{2}T,...,\hat{\beta}_{q}T\}
\end{align*}

The two games MW$_{1A}$\ and MW$_{1B}$\ are equivalent forms of the same game
and they both lead to the production of the same output state and expected
outcomes for the players.

If we do not allow the players operations $\{\hat{\alpha}_{k}\}$\ and
$\{\hat{\beta}_{k}\}$ that change the degree of entanglement then it is clear
that the resultant quantum game, a version of QPD$_{\hat{M}}$ with restricted
operator sets, is not a proper quantization of the classical game [7]. As we
have noted above, however, this game \textit{can} be simulated by playing a
game with classical coins if the outcomes are assigned probabilities in
accordance with the calculated quantum mechanical distributions over the
eigenstates, even though the resultant distributions cannot be reproduced by
any mixed strategy in the classical game with the standard assignment of
outcomes. Of course, if we allow the entanglement operator to depend on some
parameter so that varying this parameter changes the degree of entanglement
produced then the classical game can be obtained in the appropriate limit.

Entanglement often leads to a richer behaviour for a quantized system that
cannot be reproduced by the equivalent classical system. A\ playable game,
however, is not just characterized by a physical system but also by some
computation on the behaviours of that physical system. It is this computation
that determines the outcomes for the players for a particular output state.
The players have to perform this computation in order to select their best
move. Even if we keep the computation fixed for both the classical and quantum
systems, comparing the two in a game-theoretic sense is problematical because
the purpose of game theory is to determine what rational players would choose
from a set of available moves, or strategies. If we allow a larger set of
available moves in the quantum case, then it is not clear in what sense we are
playing an equivalent game.

So is there any sense in which we may observe truly quantum effects when
playing a game with quantum objects? As we have seen, the expected outcomes in
a quantum game can be reproduced by an appropriate game played with classical
coins. The choice of strategies is effectively the same in the quantum game
and its classical simulation. In general, in order to see quantum effects in
an entangled system we must look to an examination of the joint properties of
the constituent sub-systems. In order to see a violation of the Bell
inequality with 2 spin-1/2 particles, for example, we must examine the
correlations between 3 different measurements. So if we want to see
non-classical effects in quantum games it is in the statistical correlations
between the outcomes that we might look first [9-11]. We would expect there to
be a similar inequality between the correlations in the outcomes of a game
that is violated when playing with quantum objects. An examination of this
question will be published elsewhere.

\subsection{Playing QPD$_{\hat{M}}$ with Partially Entangled States}

One way to achieve these kinds of quantum games with our 2 spin-1/2 particles
is simply to input a state $\left\vert \xi\right\rangle =T(\lambda)\left\vert
0\right\rangle _{z,A}\otimes\left\vert 0\right\rangle _{z,B}$ to the game of
the previous section, where $T(\lambda)$\ is a parameter-dependent entangling
operator and we choose $\lambda$\ such that the input state is only partially
entangled. Provided the players are not allowed operations that can increase
the degree of entanglement then the resultant game will always lead to a
partially entangled output (or a completely disentangled output if we allow
interactions that can only \textit{decrease} the degree of entanglement), that
is, the output state $\left\vert \psi\right\rangle \in\Psi_{p}^{e}$ and this
is an instance of a game of type $G_{3}$.

Another way to achieve partially-entangled output states is to play a game of
the EWL-form. In this game $T$ produces maximally entangled states from the
ground state $\left\vert 0\right\rangle _{z,A}\otimes\left\vert 0\right\rangle
_{z,B}$ and the players are only allowed operations that act on the sub-spaces
independently. The state is then subject to a disentangling operation $T^{-1}$
before the final measurement is made. This version of QPD$_{\hat{M}}$ is that
considered by EWL in their original work on quantum games [4]. If we consider
the index of correlation, $I_{c}$, which is a measure of the information
content of the correlation between two systems [23,24], then for our particles
A and B, $\ I_{c}=S_{A}+S_{B}$ where $S_{A(B)}$ is the entropy of particle
A(B) given by the usual trace over the reduced density operators so that
$S_{A(B)}=-$Tr$\left(  \rho_{A(B)}\ln\rho_{A(B)}\right)  $. For the operator
$T$, which produces Bell basis states from the computational basis, $I_{c}$ is
a maximum and equal to $2\ln2$. If the players have access only to unitary
operations $U$ that act independently on the sub-spaces $H_{A}$ and $H_{B}$
then the entropies $S_{A(B)}$\ are invariant under transformations of the form
$\rho_{A(B)}^{\prime}=U\rho_{A(B)}U^{-1}$. Thus, the allowed operations of the
players cannot change the information content of the correlation and the state
remains maximally entangled after the operations of the players. The
subsequent disentangling operation can, therefore, only reduce the
entanglement of the particles. If the players have access to operations in
their available sets which, when taken together, switch between eigenstates of
the Bell basis generated by $T$ then the disentanglement is perfect and will
result in the output of a tensor product state. Thus the the output state from
this game satisfies $\left\vert \psi\right\rangle \in\Psi^{i}\cup\Psi_{p}^{e}$
and this is an instance of a game of type $G_{2}$.

As we saw in section 2, games of the EWL type can be cast in one of several
different equivalent forms. Here we are fixing the input state to be
$\left\vert \psi_{0}\right\rangle =\left\vert 0\right\rangle _{z,A}%
\otimes\left\vert 0\right\rangle _{z,B}$\ with the final measurement being
$\hat{M}=\hat{\sigma}_{z}^{A}\otimes\hat{\sigma}_{z}^{B}$\ and there being no
`imposed' final state transformation so that the output state is a result
entirely of the actions of the players. There are 3 equivalent forms of
QPD$_{\hat{M}}$\ that satisfy these conditions such that the game is of the
type $G_{2}$:\medskip

EWL$_{1}$ : input state $\left\vert \psi_{0}\right\rangle $, operator sets
$\left(  T^{-1}\hat{\Omega}_{A}T\right)  ,\left(  T^{-1}\hat{\Omega}%
_{B}T\right)  $, no final state transformation, commuting game if $\left[
\hat{\alpha}_{k},\hat{\beta}_{j}\right]  =0~~\forall j,k$.

EWL$_{2A}$ : input state $\left\vert \psi_{0}\right\rangle $, operator sets
$\left(  \hat{\Omega}_{A}T\right)  ,\left(  T^{-1}\hat{\Omega}_{B}\right)  $,
no final state transformation, player \textit{A} plays first

EWL$_{2B}$ : input state $\left\vert \psi_{0}\right\rangle $, operator sets
$\left(  T^{-1}\hat{\Omega}_{A}\right)  ,\left(  \hat{\Omega}_{B}T\right)  $,
no final state transformation, player \textit{B} plays first\medskip

The symmetric form, EWL$_{1}$,\ is the most appealing and in the version of
QPD$_{\hat{M}}$\ considered originally by EWL the available operator sets for
the players are the set of rotations on a single particle so that player
\textit{A} operates on particle A and player \textit{B} operates on particle
B. This game is then commuting since, clearly, $\left[  \hat{\alpha}_{k}%
,\hat{\beta}_{j}\right]  =0~~\forall j,k$ \ Strictly speaking, of course, the
operator choices for the players in the EWL game form a continuum. Playing the
game with the full set of available operators amounts, therefore, to being
able to select an element from an uncountably infinite set.

\subsection{Which Game Should we Play?}

Let us consider our players sitting down to play a game of QPD$_{\hat{M}}%
$\ with input state and final measurement as above and in which the physical
device that implements the game asks the players to select between the
following options before actually playing the game:\medskip

\textbf{Option 1} : the players can choose an arbitrary rotation on a single
particle such player \textit{A}(\textit{B}) operates only on particle A(B).
This leads to a game of the type $G_{1}$ and produces output states
$\left\vert \psi\right\rangle \in\Psi^{i}$.\medskip

\textbf{Option 2} : the players can choose an operator from the sets $\left(
T^{-1}\hat{\Omega}_{A}T\right)  ,\left(  T^{-1}\hat{\Omega}_{B}T\right)  $ in
which the sets $\hat{\Omega}_{A}$\ and $\hat{\Omega}_{B}$\ are those available
in option 1. This leads to a game of the form EWL$_{1}$, that is, of type
$G_{2}$ which produces output states $\left\vert \psi\right\rangle \in\Psi
^{i}\cup\Psi_{p}^{e}$.\medskip

\textbf{Option 4} : the players can choose an operator from the sets $\left(
\hat{\Omega}_{A}T\right)  $ and $\hat{\Omega}_{B}$ in which the sets
$\hat{\Omega}_{A}$\ and $\hat{\Omega}_{B}$\ are those available in option 1.
This leads to a game of the form MW$_{1A}$, that is, of type $G_{4}$ which
produces output states $\left\vert \psi\right\rangle \in\Psi_{\max}^{e}$. This
is equivalent to the game MW$_{1B}$ and so this is not listed as a separate
option.\medskip

\textbf{Option }\emph{c} : the players can select any operator from the
complete set of unitary operations on two spin-1/2 particles. This is the
complete game of QPD$_{\hat{M}}$\ and we label this as $G_{c}$. In this game
we have $\left\vert \psi\right\rangle \in\Psi$.\medskip

Which of these options, or games, would rational players choose to play?
Clearly, option\textit{\ c} would be discarded in the pure strategy case as
there is no computation that will enable the player making the first move to
choose a preferred strategy. In a sense here we have a family of games in
which the choice of which game to play is itself part of the strategy
selection of the players.

\section{Gaming the Quantum}

The approach to quantum games we have presented here focuses on the notion of
\textit{playability}, emphasizing that if we actually want to play a game in
some reality the constituent elements that implement the game are described by
physical objects and processes. By abstracting out the essential physical
elements of a playable game in this fashion we are able to develop what we
believe to be the correct perspective for drawing a correspondence between
playing games with classical and quantum objects. From this perspective,
therefore, we are approaching the notion of a quantum game not by trying to
quantize a classical game, but by considering the implications of what it
means to play a game with quantum objects and processes; in essence we might
say we are `gaming the quantum' rather than `quantizing a game'.

With this viewpoint we can see that a quantum system generates a family of
playable games for a given set of preference relations over the measurement
outcomes. The different members of this family are generated by the different
subsets of the operations of the players. Thus, QPD$_{\hat{M}}$\ represents a
family of playable games in which the players have the preference relations of
Prisoner's Dilemma over the eigenstates of $\hat{M}$. Different members of
this family can be played by restricting the operations available to the
players. The quantum games considered by Eisert, Wilkens and Lewenstein [4],
and Marinatto and Weber [8] are members of the family QPD$_{\hat{M}}$.

By shifting the perspective to that of gaming the quantum we are less
interested in questions of whether any particular classical game is contained
within the quantum version as a proper quantization in the appropriate limit,
although it is important to be mindful of such questions if we want to draw
any correct comparison between the quantum and classical versions of any
particular game [7]. It is the approach of gaming the quantum that we believe
to be potentially more fruitful in addressing the kind of questions first
raised by Meyer; will quantum games allow us to develop new techniques for
quantum information processing? Thus, if we are able to cast quantum control
problems, or quantum algorithms, or quantum circuit design in the form of a
game, can we relate the optimal outcome of the resultant game to an algorithm,
or circuit, that is in some sense optimal? This is the direction we have taken
elsewhere, where the geometric picture of considering the output state to be
the result of some competition between optimal strategy choices has been found
to be useful [21,22].

In the perspective we have developed the quantization of the physical
implementation of an instance of a game becomes clear. To what extent this
results in something we might call a quantized game, however, is less clear.
As we have seen, if we are only interested in the expected outcomes and how
they impact upon the strategy choices of the players, then any playable
quantum game can be reproduced by \textit{some} classical game played entirely
with classical coins, provided we can compute the probabilities for the
outcomes for any given output state.

\bigskip

\bigskip

\textbf{References}

\begin{enumerate}
\item K. Binmore, \textit{Playing for Real: A Text on Game Theory}, OUP USA, (2007)

\item J.A. Wheeler and W.H. Zurek (eds.), \textit{Quantum Theory and
Measurement}, Princeton University Press, (1983)

\item D. Meyer, \textit{Quantum Strategies, }Phys. Rev. Lett., \textbf{82},
1052-1055, (1999)

\item J. Eisert, M. Wilkens and M. Lewenstein, \textit{Quantum Games and
Quantum Strategies}, Phys. Rev. Lett., \textbf{83}, 3077-3080, (1999)

\item D. Deutsch, \textit{Quantum theory, the Church-Turing principle and the
universal quantum computer}, Proc. Roy. Soc. A, Mathematical and Physical
Sciences \textbf{400} (1818), 97--117, (1985)

\item P. W. Shor, \textit{Polynomial-Time Algorithms for Prime Factorization
and Discrete Logarithms on a Quantum Computer}, SIAM J. Comput., \textbf{26}
(5), 1484--1509, (1997)

\item S. A. Bleiler, \textit{A Formalism for Quantum Games I - Quantizing
Mixtures}, Portland State University, preprint, available at
http://arxiv.org/abs/0808.1389, (2008)

\item L. Marinatto and T. Weber, \textit{A quantum approach to static games of
complete information}, Phys. Lett. A, \textbf{272}, 291-303, (2000)

\item A. Iqbal and D. Abbott, \textit{Constructing quantum games from a system
of Bell's inequalities}, Phys. Lett. A, \textbf{374}/31-32, 3155-3163, (2010)

\item A. Iqbal\textit{, Playing games with EPR-type experiments}, J. Phys. A,
\textbf{38}/43, 9551-9564, (2005)

\item A. Iqbal and S. Weigert, \textit{Quantum correlation games}, J. Phys. A,
\textbf{37}, 5873-5885, (2004)

\item R. Cleve, P. Hoyer, B. Toner and J. Watrous, \textit{Consequences and
Limits of Nonlocal Strategies}, preprint, available at http://arxiv.org/abs/quant-ph/0404076v2

\item J. Shimamura, \c{S}. K. \"{O}zdemir, F. Morikoshi and N. Imoto,
\textit{Entangled states that cannot reproduce original classical games in
their quantum version}, Phys. Lett \ A, \textbf{328}, 20-25, (2004)

\item G. Brassard, R. Cleve, and A. Tapp. \textit{Cost of exactly simulating
quantum entanglement with classical communication}, Phys. Rev. Lett.,
\textbf{83}(9):1874--1877, (1999)

\item R. Renner and S. Wolf, \textit{Towards Characterizing the Non-Locality
of Entangled Quantum States}, preprint, available at
http://arxiv.org/abs/quant-ph/0211019, (2002)

\item S. J. van Enk and R. Pike, \textit{Classical Rules in Quantum Games},
Phys. Rev. A, \textbf{66,} 024306, (2002)

\item S. E. Landsburg, \textit{Nash Equilibria in Quantum Games}, Proc. Am.
Math Soc., \textbf{139}, 4423-4434, (2011)

\item G. Dahl and S. Landsburg, \textit{Quantum Strategies in Noncooperative
Games}, preprint, available at http://www.landsburg.com/dahlcurrent.pdf

\item J. Nash, \textit{Equilibrium points in n-person games}, Proceedings of
the National Academy of Sciences \textbf{36}(1), 48-49, (1950)

\item R. Penrose, \textit{Shadows Of The Mind: A Search for the Missing
Science of Consciousness}, Vintage, (2005)

\item F. S. Khan and S. J. D. Phoenix, \textit{Nash equilibrium in quantum
superpositions}, Proceedings of SPIE, \textbf{8057,} 80570K-1

\item F. S. Khan and S. J. D. Phoenix, \textit{Gaming the Quantum}, preprint,
available at http://arxiv.org/abs/1202.1142, (2012)

\item S. M. Barnett and S. J. D. Phoenix,\textit{\ Information Theory,
Squeezing and Quantum Correlations}, Phys. Rev. A, \textbf{44}, 535, (1991)

\item S. M. Barnett and S. J. D. Phoenix, \textit{Information-Theoretic Limits
to Quantum Cryptography}, Phys. Rev. A, \textbf{48,} R5, (1993)
\end{enumerate}

\end{document}